\documentclass[usenatbib]{mnras}

\usepackage{newtxtext,newtxmath}
\usepackage{lineno}

\usepackage[T1]{fontenc}

\DeclareRobustCommand{\VAN}[3]{#2}
\let\VANthebibliography\thebibliography
\def\thebibliography{\DeclareRobustCommand{\VAN}[3]{##3}\VANthebibliography}

\usepackage{comment}
\usepackage{graphicx}
\usepackage{amsmath}
\usepackage{mathtools}
\usepackage{bm}
\usepackage[dvipsnames]{xcolor}
\usepackage{gensymb}
\usepackage{multirow}
\usepackage{bigdelim}

\newcommand{\dash}{\text{--}}
\newcommand{\xhat}{\hat{\bm{x}}}
\newcommand{\yhat}{\hat{\bm{y}}}
\newcommand\timesdiv{\mathbin{\vcenter{\hbox{%
   $\begin{array}{@{}c@{}}\times\\[-1.667ex]\div\end{array}$}}}}
   
\title[On the spatial correlations of photospheric magnetic turbulence]{von K\'arm\'an--Howarth Similarity of Spatial Correlations and the Distribution of Correlation Lengths in Solar Photospheric Turbulence}

\author[R. Chhiber et al.]{
Rohit Chhiber$^{1,2}$\thanks{E-mail: rohit.chhiber@nasa.gov},
Raphael Attie$^{3,2}$,
William H. Matthaeus$^{1}$,
Sohom Roy$^{4,1}$,
and Barbara J. Thompson$^2$
\\ \\
$^{1}$Department of Physics and Astronomy, University of Delaware, Newark, DE 19716, USA\\
$^{2}$Code 671, Heliophysics Science Division, NASA Goddard Space Flight Center, Greenbelt, MD 20771, USA\\
$^{3}$George Mason University, Fairfax, VA 22030, USA\\
$^{4}$Space Research Institute, Austrian Academy of Sciences, Schmiedlstraße 6, 8042, Graz, Austria
}

\date{Accepted XXX. Received YYY; in original form ZZZ}

\pubyear{2025}

\begin{document}
\label{firstpage}
\pagerange{\pageref{firstpage}--\pageref{lastpage}}
\maketitle

\begin{abstract}
Fluctuations in the Sun's photospheric magnetic field are the primary source of the turbulence that can heat and accelerate the solar atmosphere, and thus play an important role in the production and evolution of the solar wind that permeates the heliosphere. A key parameter that characterizes this turbulence is the correlation scale of fluctuations, which determines the injection of turbulent energy into the plasma and the diffusive transport of solar energetic particles. This study employs magnetogram data from the Helioseismic and Magnetic Imager on the Solar Dynamics Observatory to characterize an ensemble of spatial autocorrelation functions (ACFs) of turbulence in the photosphere. It is shown that the two-point ACFs satisfy the similarity-decay hypothesis of von K\'arm\'an and Howarth, a fundamental property of turbulent systems: rescaling the ACFs by their respective energies and correlation lengths yields a quasi-universal exponential form. The probability distribution function of transverse correlation lengths (\(\lambda\)) is shown to be approximately log-normal, which is consistent with observations of turbulence in the solar wind. A ``mosaic'' of the spatial distribution of \(\lambda\) over the photosphere is presented; the ``quiet Sun'' tends to have \(\lambda\sim 1500\) km (albeit with a wide distribution), which is close to the scale of solar granulation; systematically longer lengths are associated with active regions. A positive correlation is observed between mean magnetic field magnitude and \(\lambda\), and empirical fits quantify this relationship. These results improve our understanding of solar turbulence while providing observational constraints for models that describe turbulence transport from solar and stellar photospheres into their atmospheres.
\end{abstract}



\section{Introduction}\label{sec-Intro}

It is widely regarded that that the solar plasma is highly dynamic and probably turbulent in a broad sense \citep{Miesch2005LRSP}. The solar dynamo is an example of thermally-driven turbulent convection, likely magnetohydrodynamic in view of the influence of locally strong magnetic fields. Above the photosphere (which will be the focus of the present study), the chromosphere is collisional and dynamic with widely varying temperatures and many ionization states \citep{Carlsson2019ARAA}. Around the transition region the plasma becomes collisionless, and it is widely believed that turbulence plays a central role in the heating of the coronal plasma and the acceleration of the solar wind \citep{bruno2013LRSP,deforest2018ApJ,Rivera2024Sci}. 
In contrast, turbulence in the solar photosphere has received relatively limited attention \citep[see][for reviews]{Petrovay2001SSR,Rincon2018LRSP}. When measurements become available, a frequent first step has been to compute spectra, to evaluate basic turbulence ideas such as Kolmogorov's inertial range and its variations \citep{pope2000book,Kiyani2015RSPTA}. Here we take a step back from that practice and test ideas that underlie spectral theories, as well as theories of variability of basic turbulence parameters.

Our focus will be on the properties of two-point spatial autocorrelation functions (ACFs) of magnetic field fluctuations in the photosphere. These ACFs occupy a central position in turbulence studies, since they are formally related to the (more commonly evaluated) turbulent energy spectrum, and enable the computation of the correlation scale of turbulence \citep[e.g.,][]{matthaeus1982JGR}. The hypothesis that the ACF has a quasi-universal form \citep{karman1938prsl} despite the large degree of variability typically seen in turbulent flows \citep{oboukhov1962JFM} is a stepping stone towards modeling approaches that are widely used in heliophysics \citep[e.g.,][]{hossain1995PhFl,matthaeus1999ApJL523}. Besides the formal significance of the ACF, a practical consideration is the need to constrain the value of the  correlation length in the photosphere. This parameter determines the scale (and rate) of turbulent energy injection, and therefore its distribution over the photosphere is an important boundary condition for coronal heating models \citep[e.g.,][]{cranmer2007ApJS,verdini2007apj}. The probability distribution of correlation lengths relates to the generation of ``\(1/f\) noise'' that is observed in magnetic power spectra in the photosphere and in interplanetary space \citep[e.g.,][]{Wang2024SoPh}. The implications extend to space weather, since the diffusive spreading of magnetic field lines and solar energetic particles (SEPs) during their transport through the heliosphere sensitively depends on 
the correlation length \citep[e.g.,][]{chhiber2017ApJS230,chhiber2021ApJ_flrw,Engelbrecht2022SSR}.

While the properties of turbulent ACFs and correlation lengths in interplanetary space (helioradii greater than \(\gtrsim0.3\) AU) have been studied for decades \citep[e.g.,][]{matthaeus1982JGR,Ruiz2014SoPh,Roy2021ApJ}, recent years have seen an increased interest in their behavior closer to the Sun, observed in-situ  \citep{cuesta2022ApJL} and remotely \citep[][]{Sharma2023NatAst,Bailey2025ApJ,Hahn2025ApJ_corrlen}. These properties have also been examined in the photosphere, specifically within a coronal hole \citep{Abramenko2013ApJ} and active regions \citep{Abramenko2024SoPh}.
As NASA's Parker Solar Probe continues to provide in-situ observations of  the corona at heliodistances near \(10~R_\odot\) \citep{fox2016SSR}, we have an opportunity to obtain a more comprehensive and panoramic view of the global distribution and evolution of the ACFs and associated correlation scales, from the photosphere to the corona and beyond.

In this work we present an initial study of some key properties of these correlation functions in the photosphere, employing data from a typical full-disk magnetogram observed around solar minimum. 
An organizing concept is the \cite{karman1938prsl}
{\it self-preservation hypothesis}
which posits that the dynamics the ACFs 
follow a two parameter similarity 
law during decay. Notably 
the two parameters are the energy density and a similarity scale, the latter usually taken to be the correlation scale.
This highlights the significance of the correlation scale in relation 
to the ACFs, and consequently also 
in relation to the spectra and its many applications.
A more detailed background of relevant theory and observations is given in Sec. \ref{sec:backg}, with a description of the data used in Sec. \ref{sec:Data}. Results are presented in Sec. \ref{sec:res}, followed by a summary and discussion in Sec. \ref{sec:Disc}. Appendix \ref{sec:app0} contains some particulars of the ACF calculation, Appendix \ref{sec:app1} relates the correlation length to the energy spectrum, and Appendix \ref{sec:app2} contains supplemental analyses of ACFs.

\section{Theoretical and Observational Background}\label{sec:backg}

The two-point, single-time spatial autocorrelation function (ACF) for a turbulent field \(\bm{b}\) is defined as the \(2^\text{nd}\)-order tensor \citep[e.g.,][]{pope2000book}
\begin{equation}
    R_{ij}(\bm{\ell},t) = \langle b_i(\bm{r},t) ~b_j(\bm{r}+\bm{\ell},t) \rangle, \label{eq:R0}
\end{equation}
where \(t\) is the time coordinate, subscripts \(i\) and \(j\) refer to vector components in a three-dimensional (3D) coordinate system, \(b_i(\bm{r},t)\) is the field's \(i\) component at a position \(\bm{r}\), and \(b_j(\bm{r}+\bm{\ell},t)\) is the field's \(j\) component at a position ``lagged'' by a vector displacement \(\bm{\ell}\) relative to \(\bm{r}\). The \(\langle \dots \rangle \) operator indicates an average over a suitably-defined statistical ensemble. Assumption of statistical homogeneity in space implies that the ACF depends only on the spatial lag \(\bm{\ell}\) and is independent of \(\bm{r}\) \citep{Batchelor1953book}. Note that the trace of the tensor with \(\ell=0\) is (twice) the average energy per-unit-mass of the turbulent field (with the magnetic field $b$ in Alfv\'en speed units). 
One can specialize to the cases where the field's vector components are either parallel or perpendicular to \(\bm{r}\), thus defining the \textit{longitudinal} and \textit{transverse} ACFs, respectively,
in the standard hydrodynamic nomenclature \citep{Batchelor1953book}. We recall that in plasma physics and MHD, spectra and correlation functions 
are often anisotropic relative to the magnetic field direction, and in such cases one may define 
{\it parallel} and {\it perpendicular}
correlations relative to that distinct direction \citep[e.g.,][]{oughton2015philtran}.

The characteristic scale up to which the fluctuations may be considered correlated, or the \textit{correlation length}, is defined to be \citep{Batchelor1953book,matthaeus1982JGR}
\begin{equation}
    \lambda= \int_0^\infty d\ell~ R(\ell)/R(0), \label{eq:lambda1}
\end{equation}
where we have suppressed the time coordinate and the tensor subscripts on \(R\) and picked a specified direction for the lag, which is expressed as the scalar \(\ell\). An alternative estimate for the correlation scale is provided by the ``\(e\)-folding'' length, i.e., the length where \(R(\ell)/R(0)=1/e=0.3678...\) \citep[e.g.,][]{Roy2021ApJ}. \(\lambda\) is usually interpreted as the scale of the largest turbulent structures (``eddies''), at which energy from large-scale ``driving'' is injected into the turbulent energy cascade. Relatedly, the correlation length is associated with the low-wavenumber end of the inertial-range energy spectrum, often called the bend-over scale \citep{Kiyani2015RSPTA}.\footnote{See also Appendix \ref{sec:app1}.} 

\cite{karman1938prsl} derived a dynamical equation for the evolution of the \(2^\text{nd}\)-order correlation tensor for homogeneous isotropic turbulence, and studied its solution in the case when the shape of this tensor remains similar and only its scale changes. In other words, the correlation functions given by this solution are ``self preserving'' in the sense that their form remains the same at all instants even as their characteristic length scale varies (for scales large compared to the viscous or dissipation scale). Under such conditions, it was found that the nonlinear terms in the dynamical equation for turbulent energy decay can be effectively modeled in terms of characteristic length and time scales, thus enabling the use of so-called \textit{closure models} in lieu of complete analytical solutions. For the hydrodynamic case the von K\'arm\'an decay laws are \(\frac{du^2}{dt}=-\alpha u^3/\lambda\) and \(\frac{d\lambda}{dt}=\beta u\), where \(u\) is the characteristic turbulent speed, and \(\alpha\) and \(\beta\) are constants. This type of closure modeling is extensively used in computational and theoretical study of turbulent flows in natural and engineering systems \citep[e.g.,][]{pope2000book}, including models that describe the transport of turbulence in the corona and the solar wind \citep{zhou1990transport,tu1995SSRv,zank1996evolution,matthaeus1999ApJL523,breech2008turbulence,verdini2010ApJ,vanderholst2014ApJ,Lionello2014ApJ,usmanov2018,zank2018ApJ}. The von K\'arm\'an--Howarth (henceforth vK-H) similarity hypothesis for ACFs therefore provides a formal mathematical basis for models that employ turbulent dissipation as a mechanism to heat and energize plasmas in the heliosphere.

To be specific, the vK-H similarity hypothesis asserts that the functional form of the ACF \(R(\ell,t)\) is self-preserving in the sense that it can be expressed at any instant of the turbulent decay as 
\begin{equation}
    R(\ell,t) = u^2(t) \mathscr{R}[\ell/\lambda(t)]. \label{eq:vkH}
\end{equation}
Here \(\mathscr{R}\) is a universal function that describes the dynamics of the correlation function in the intermediate range of scales that is much larger than the dissipative scales, and smaller than any specific large, coherent structures (like active regions in the photosphere or stream-interaction regions in the solar wind) that might introduce non-homogeneity. The experimental verification of this hypothesis remains a cornerstone of hydrodynamic turbulence theory \citep{Batchelor1948RSPSA,Stewart1951RSPTA}.

While the turbulent correlation tensor was originally defined in terms of the velocity field in hydrodynamic systems, studies of space and heliospheric turbulence have focused on the magnetic field due to its readily available measurements. ACFs, correlation lengths, and power spectra of the magnetic field in the solar wind have been extensively studied using observations, theory, and numerical simulations \citep[][and references therein]{bruno2013LRSP}. It was only recently, however, that the vK-H similarity hypothesis was directly validated for a turbulent astrophysical system for the first time, using in-situ observations of magnetic and plasma fields in near-Earth space \citep{Roy2021ApJ,Roy2022PRE}. 

In the context of astrophysical systems like the solar wind, the correlation length [Eq. \eqref{eq:lambda1}] also plays a key role in models of turbulent transport and heating alluded to above, as well as in the transport and scattering of SEPs and galactic cosmic rays (GCRs) \citep[][and references therein]{Engelbrecht2022SSR}. Its evolution with distance from the Sun and its variability in the complex and dynamic solar-wind environment are therefore of considerable relevance to space weather modeling and prediction \citep[e.g.,][]{Whitman2023AdSpR}. Until recently, correlation lengths of turbulent magnetic fields in the heliosphere had mainly been studied using in-situ observations of interplanetary solar wind, at distances above \(0.3\) AU from the Sun \citep[e.g.,][]{Ruiz2014SoPh,matthaeus1982JGR,smith2001JGR,adhikari2017ApJ}. These studies have found a systematic increasing trend in \(\lambda\) with distance from the Sun, which is attributed to (i) turbulence becoming more well-developed as the solar wind flows outward \citep[with larger spatial scales getting involved in the inertial-range cascade;][]{matthaeus1998JGR,bruno2013LRSP,deforest2016ApJ828,Chhiber2018PhDT}, and (ii) geometrical expansion of magnetic structures like flux tubes \citep{Hollweg1986JGR}. NASA's Parker Solar Probe mission \citep[PSP;][]{fox2016SSR} has extended the availability of in-situ observations to distances of \(\sim0.05\) AU and recent studies have employed these data to investigate \(\lambda\) in the ``young solar wind'' \citep{chhiber2021ApJ_psp,cuesta2022ApJL}. One finds an increase in \(\lambda\) from \(\sim 10^4\) km near 0.1 AU to \(\sim 10^6\) km near Earth, with a further increase to around \(10^7\) km by 40 AU (see Table \ref{table:1}).

Of course, the characteristic scales of turbulence in the solar wind are, to a large extent, determined by the corresponding scales at the Sun. One must rely solely on remote sensing of the photosphere, chromosphere, and the low corona in order to obtain observational constraints on these values. Perhaps the earliest observational analysis of turbulent ACFs in the magnetic photosphere was carried out by \cite{Abramenko2013ApJ}, who employed magnetograms to estimate \(\lambda \sim 10^3\) km at the base of a coronal hole. A very recent study by \cite{Abramenko2024SoPh} found much larger values (\(\sim 10^4\) km) associated with solar active regions (ARs). The last couple of years have also seen studies of correlation lengths evaluated just above the photosphere using remote observations off the solar limb \citep{Sharma2023NatAst,Bailey2025ApJ,Hahn2025ApJ_corrlen}, finding values of around \(1500\dash 5000\) km in the chromosphere that increase across the transition region to around \(3500\dash 8000\) km at the base of the corona (see Table \ref{table:1}). Knowledge of the correlation lengths in these regions provides a crucial boundary condition for solar wind models with turbulent heating, and is also of relevance to SEP injection and transport.

Like most properties of a turbulent system, the probability density function (PDF) of correlation lengths tends to show a large variance. Observational studies of this PDF in interplanetary space suggest a lognormal form, i.e., \(\log(\lambda)\) has a normal distribution \citep{Ruiz2014SoPh,isaacs2015JGR120,Pradata2025ApJ}. Lognormal distributions are widely observed in the physical, biological, and social sciences; they are associated with  systems containing a large number of independent multiplicative (non-linear) effects (in contrast to the additive effects that produce normal distributions), leading to asymmetric PDFs with a long tail towards large values higher than the mean \citep[e.g.,][]{Limpert2001Biosci}. The lognormality of correlation scales has been linked to the generation of ``\(1/f\)'' noise in the low-frequency (\(f\)) spectrum of the heliospheric magnetic field \citep[][and references within]{Wang2024SoPh}.


%
%

Our goal in the present work is to further our understanding of these inter-related topics by analyses of high-resolution magnetogram data that cover the solar disk. ACFs computed within local averaging domains over the photosphere will provide an ensemble that will be used to further constrain the values of \(\lambda\) and examine their spatial distribution. We will evaluate the vK-H similarity hypothesis for the first time in solar turbulence, and show that the PDF of photospheric correlation lengths is well-approximated by a lognormal form. We proceed by describing the magnetogram data employed in the study, below.

\section{Data}\label{sec:Data}

\begin{figure}
\centering
\includegraphics[width=0.37\textwidth]{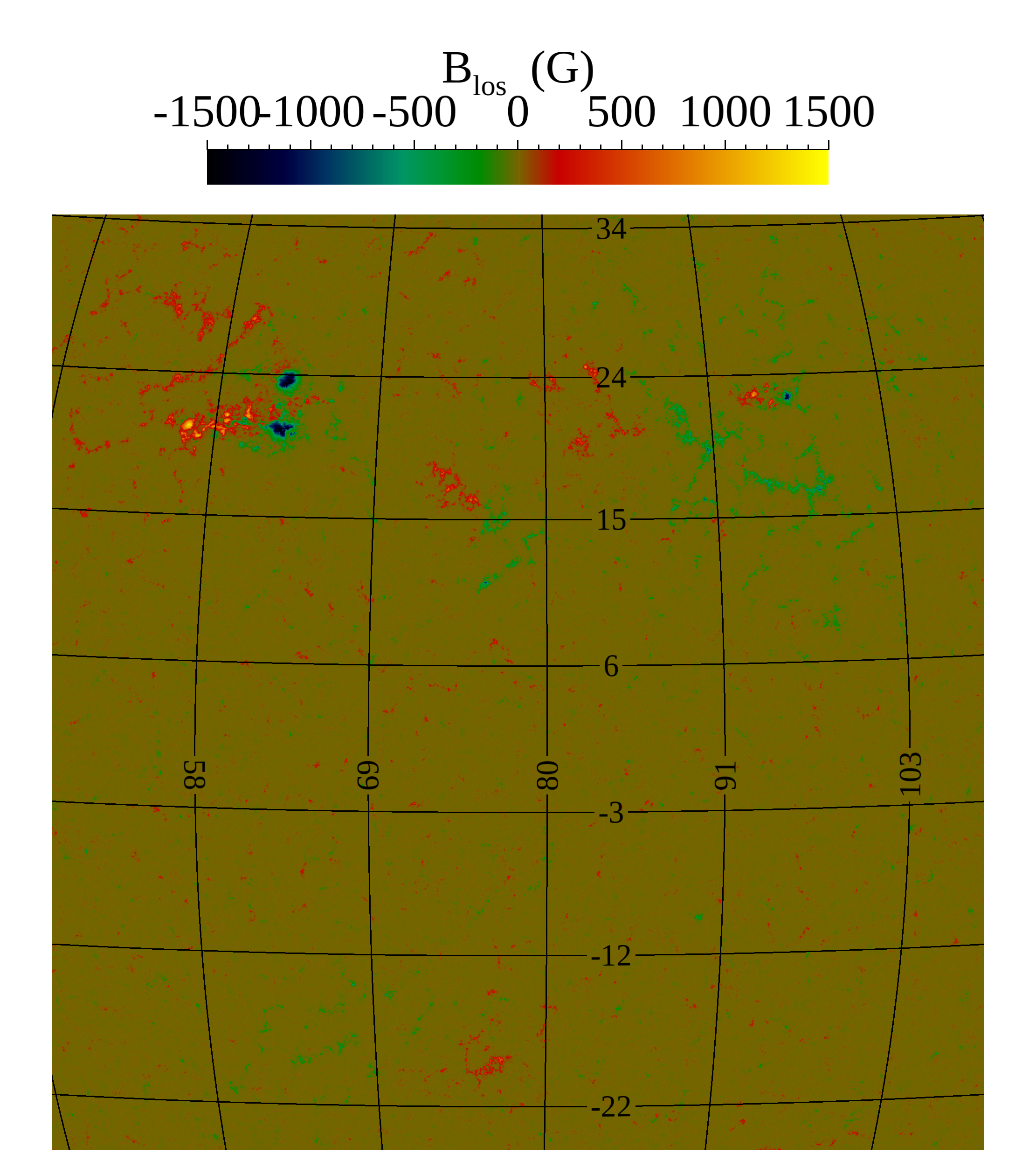}
\includegraphics[width=0.45\textwidth]{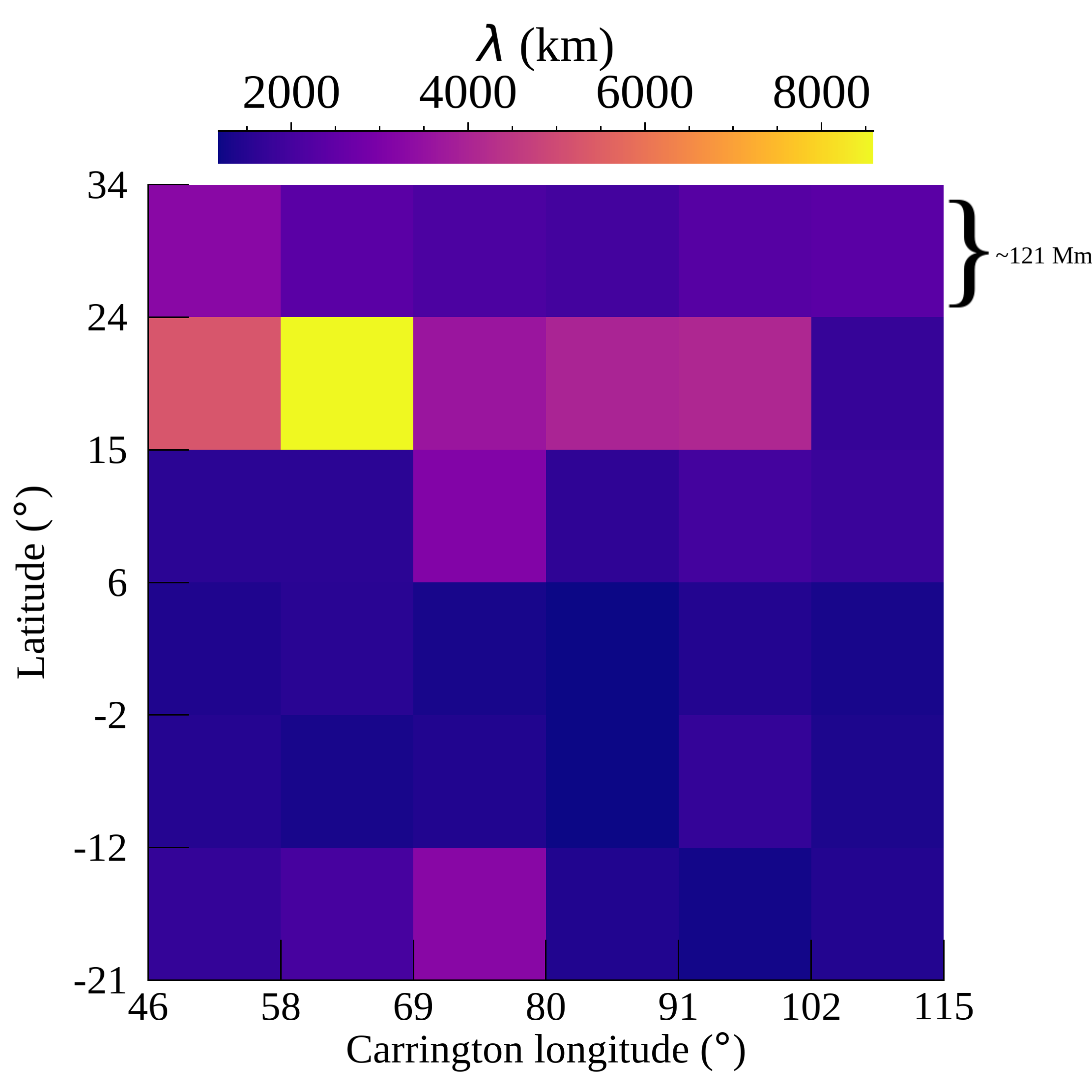}
        \caption{\textit{Top}: Line-of-sight magnetic field map, cropped at disk center from a full-disk 720-s averaged HMI magnetogram dated 2010.10.24 12:00:00 (TIA). Contours of Carrington longitude and latitude are shown. An upper limit of 1500 G has been imposed on the color map. \textit{Bottom}: Mosaic of correlation lengths of magnetic fluctuations, corresponding to the magnetic field shown in top panel. See Sec. \ref{sec:res} and App. \ref{sec:app0} for details of computation and discussion. The shown latitude/longitude tick marks are approximate. Each element (or pixel) comprising the mosaic has an approximately 121 Mm long side.}
        \label{fig:magnetogram}
\end{figure}
We analyze a full-disk magnetogram dated 2010.10.24 12:00:00 (TAI), obtained by the Helioseismic and Magnetic Imager (HMI) on board NASA's Solar Dynamics Observatory \citep{Scherrer2012SoPh}. The magnetogram provides a \(4096\times 4096\) pixel image of the line of sight (LoS) component of the magnetic-field on the full disk of the photosphere, computed as a temporal average over 720 s. The coordinate grid of solar latitude and Carrington longitude associated with each image pixel is obtained using SolarSoft world coordinate system (WCS) routines for the IDL programming language \citep{Freeland1998SoPh,Thompson2006AA}. To mitigate projection effects and the associated irregularity of the spherical coordinate grid we crop a section of the image at disk center, spanning around \(\pm 30\degree\) about the disk center. The resulting image is shown in the top panel of Fig. \ref{fig:magnetogram}, and contains \(1915\times 1921\) pixels (horizontal \(\times\) vertical) that cover both active region (AR) and quiet Sun (QS) areas.

\section{Results}\label{sec:res}


\textit{I. Computation of ACFs.} We first describe the computation of ACFs of magnetic fluctuations \(b\). From Eq. \eqref{eq:R0}, suppressing the time coordinate, we have the quantity
\begin{equation}
    R(\bm{\ell}) = \langle b(\bm{r}) b(\bm{r}+\bm{\ell}) \rangle, \label{eq:R1}
\end{equation}
which is operationally computed using the equation
\begin{equation}
    R(\bm{\ell}) = \langle B(\bm{r}) B(\bm{r}+\bm{\ell}) \rangle - \langle B(\bm{r})\rangle \langle B(\bm{r}+\bm{\ell})\rangle, 
    \label{eq:R2}
\end{equation}
where \(B(\bm{r})\) refers to the (LoS) magnetic field at a position \(\bm{r}\) on the image and \(B(\bm{r+\ell})\) is the corresponding field at a position ``lagged'' by a vector displacement \(\bm{\ell}\) relative to \(\bm{r}\). The \(\langle \dots \rangle \) operator indicates an average over suitably-defined area on the image \citep[see below; see also][for more details]{Roy2021ApJ}. Assumption of statistical homogeneity within this area implies that the ACF depends only on the spatial lag \(\bm{\ell}\) and is independent of \(\bm{r}\) \citep{Batchelor1953book,matthaeus1982JGR}. It can be seen that the above equation reduces to Eq. \eqref{eq:R1} by using the decomposition \(B=\langle B\rangle + b\) \citep[note that \(\langle b\rangle=0\);][]{Tennekes1972book}. Note that we perform our analyses for a magnetogram at a fixed instant in time (Sec. \ref{sec:Data}), allowing us to suppress the time coordinate seen in Eq. \eqref{eq:R0}; however, for our purpose the magnetogram can be considered to be a fairly typical one for solar-activity minimum conditions, and we do not expect the results to change significantly when considering other similar magnetograms.

We consider lags \(\ell_x\) and \(\ell_y\) in \(\hat{\bm{x}}\) and \(\hat{\bm{y}}\) directions, corresponding respectively to horizontal and vertical directions in the image plane. The latitude-longitude coordinate grid is irregular, and care must be taken in converting lags in pixel space to physical space; our approach is the following. We compute \(R(\ell_x)\) within rectangular subsections of the image with dimension \(200\times 40\) pixels, and \(R(\ell_y)\) within rectangular subsections of dimension \(40\times 200\) pixels. Within these averaging domains the inter-pixel spacing is relatively uniform and, apart from domains containing ARs, the distribution of magnetic field appears to follow statistical homogeneity. The inter-pixel spacing in the \(\xhat\) direction is computed for each such domain as \(\Delta_x\equiv\langle |\delta_x\phi| \rangle R_\odot \), where \(\delta_x\phi\) is the angular separation in the \(\xhat\) direction between pairs of neighboring pixels (that are adjacent in the \(\xhat\) direction), and \(\langle |\delta_x\phi| \rangle\) is the average over all such pairs within the respective domains. This mean angular separation is expressed in radians and multiplied with a solar radius \(R_\odot\) to obtain the mean inter-pixel spacing \(\Delta_x\) within a domain, for computation of \(R (\ell_x)\). A similar procedure is used to compute the mean inter-pixel spacing in the \(\yhat\) direction [\(\Delta_y\equiv \langle|\delta_y\theta|\rangle R_\odot\), where \(\theta\) indicates angular separation in the \(\yhat\) direction] within each domain, as defined above for computation of \(R(\ell_y)\).

The inter-pixel spacings \(\Delta_x\), as computed above, have a mean of 398 km and a standard deviation of 27 km, with minimum and maximum values of 364 km and 473 km, respectively; similar values hold for \(\Delta_y\). Spatial lags can then be computed as \(\ell_{i,N} = N\Delta_i\), where \(i\) indicates either \(x\) or \(y\) direction, and \(N\) is the lag expressed in number of pixels. Values taken by \(N\) range from \(0\) to one-fifth of the dimension of an averaging domain in the direction of the lag; i.e., the maximum lag is \(1/5^\text{th}\) of the ``length'' of the data record within a domain \citep[][]{matthaeus1982JGR}. ACFs are computed within each domain using Eq. \eqref{eq:R2} (see Appendix \ref{sec:app0} for further details) and normalized to their respective zero-lag values (i.e., by the mean turbulent energy of the magnetic field), thus producing ACFs with a zero-lag value of unity: \(R'(\ell_x) = R(\ell_x)/R(0)\). Note that the LoS component of \(B\) is near-radial (in a heliocentric coordinate system) and the direction of the lags is near-transverse to the radial, so we are measuring the transverse correlation function as opposed to the longitudinal one \citep{Batchelor1953book,dasso2005ApJ}.\footnote{We wish to address a subtle point regarding the interpretation of the computed ACFs as being quasi-transverse relative to the radial direction. The angle (\(\varphi\)) between the LoS and radial directions is negligible at disc center, but is \(\sim30\degree~\) at the edges of the image, which calls into question the assumption that the LoS field is near-radial here. However, since we are computing the ACF within local averaging domains that span \(\sim 6\degree\)~ in angular width, the \(\cos \varphi\)~ ``correction'' that converts the LoS component to the radial one applies as a multiplicative factor to \(B_\text{los}\) that is near-constant within each domain. The ACFs are normalized by their zero-lag value in our analysis, and therefore the above correction factor cancels out and does not affect the results in a significant way.} 

\begin{figure*}  
\centering
\includegraphics[width=0.4\textwidth]{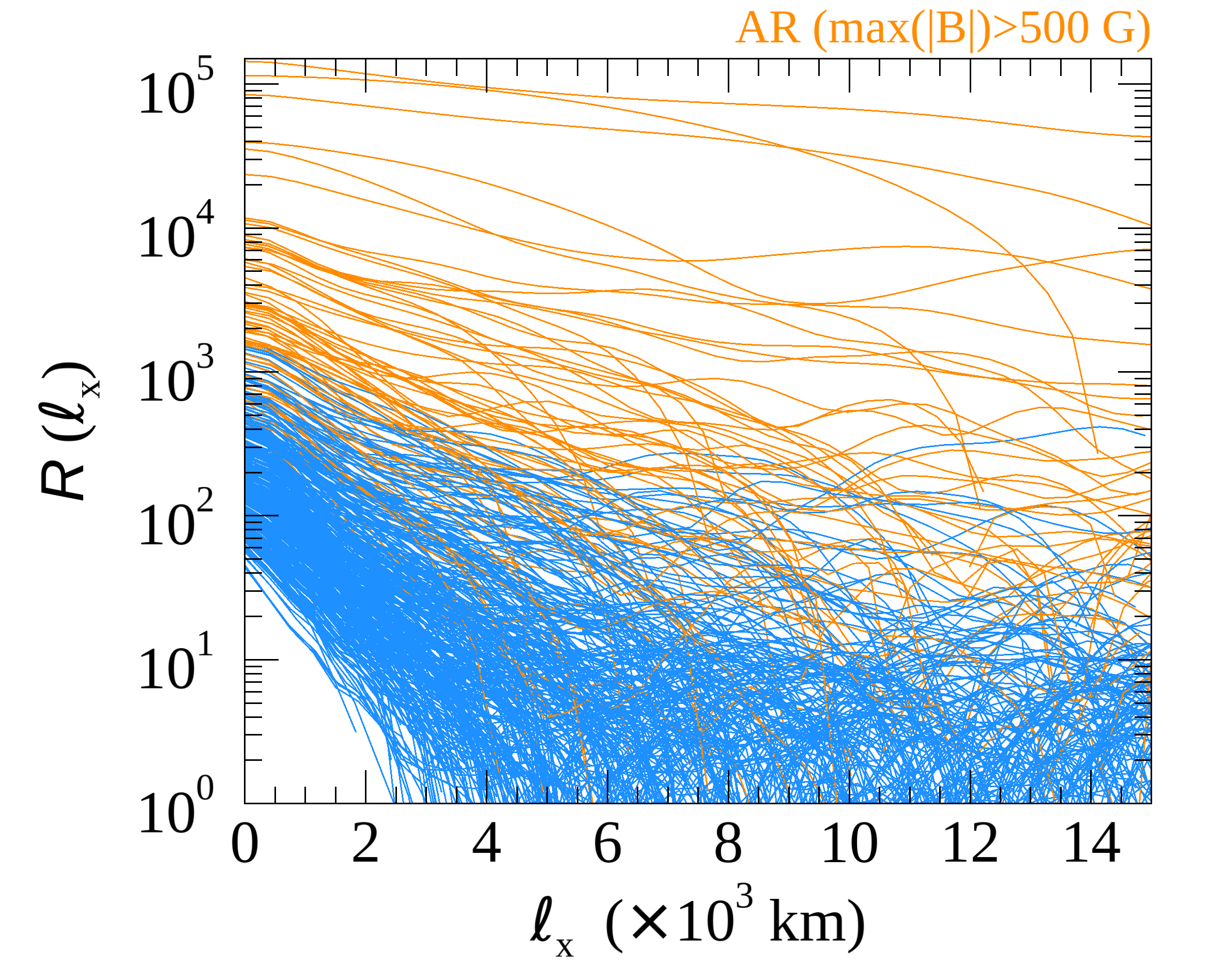}
\includegraphics[width=0.4\textwidth]{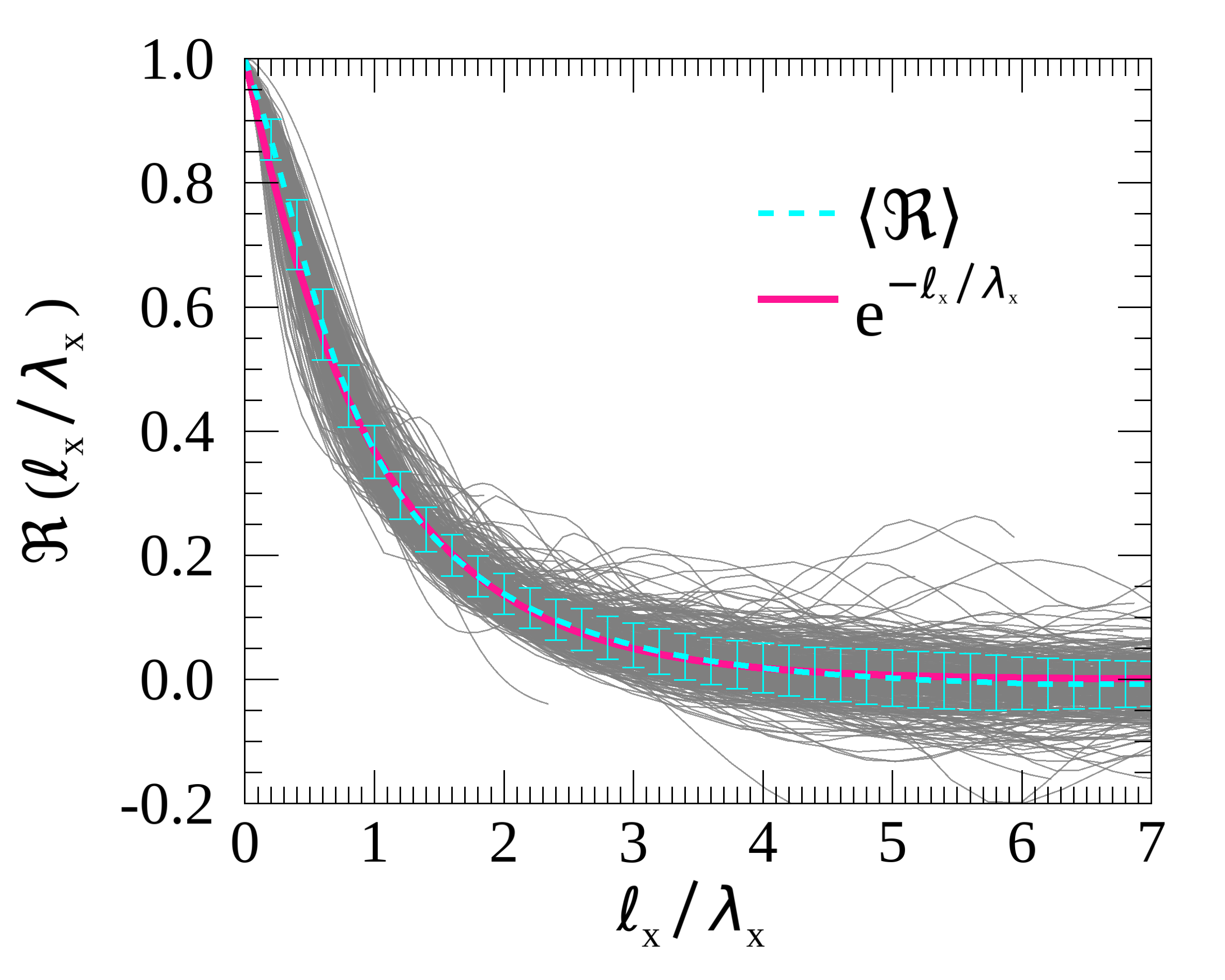}
\includegraphics[width=0.4\textwidth]{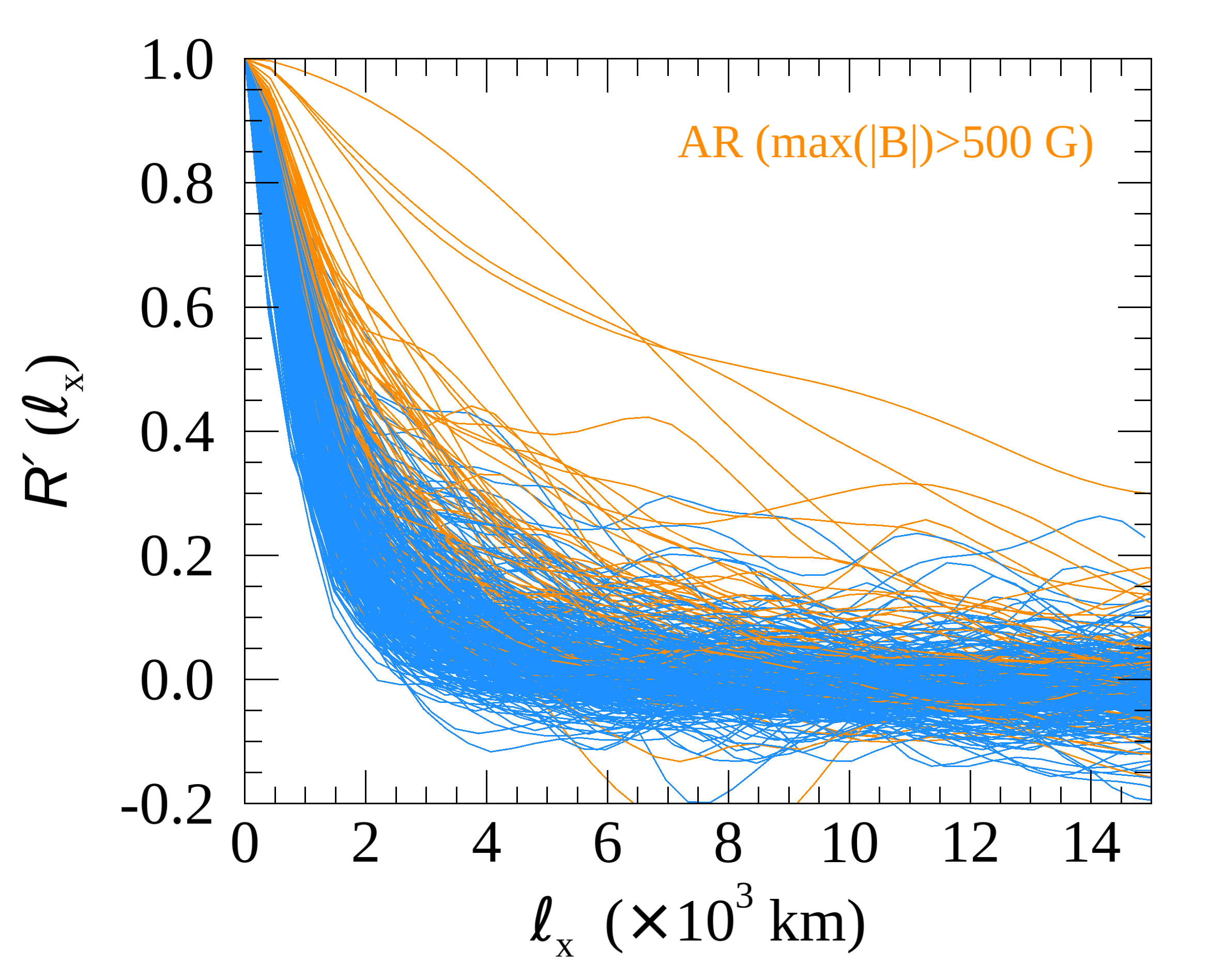}
\includegraphics[width=0.4\textwidth]{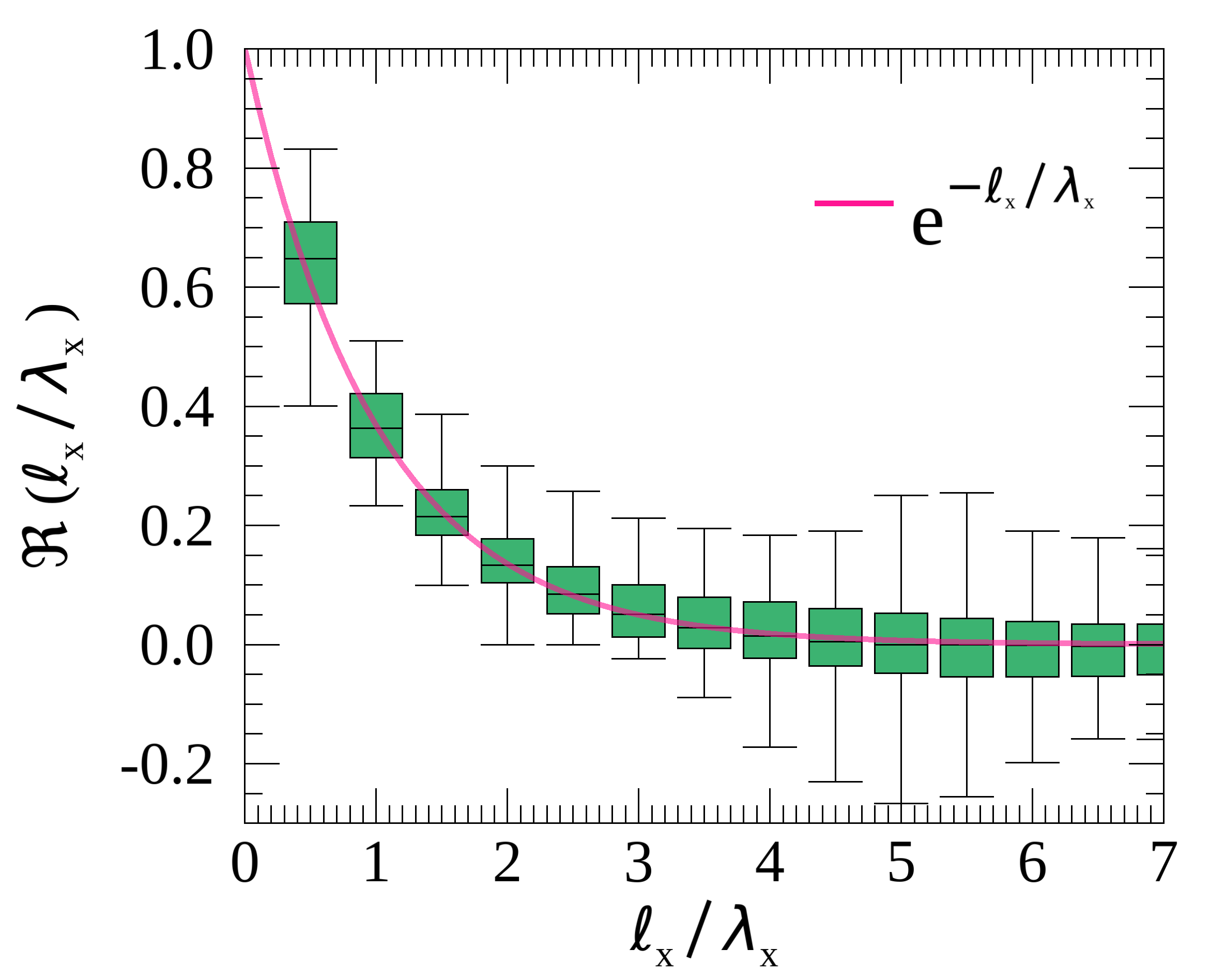}
        \caption{\textit{Top left}: \(R(\ell_x)\) are the two-point autocorrelation functions (ACFs) of magnetic fluctuations, with spatial lags \(\ell_x\) in the horizontal (\(\xhat\)) direction. \textit{Bottom left}: Each ACF is normalized by its value at zero lag: \(R'(\ell_x)=R(\ell_x)/R(0)\). \textit{Top right}: The lags associated with each ACF \(R'\) are rescaled by the respective ACF's  correlation length \(\lambda_x\). Resulting ACFs (\(\mathscr{R}\)) are plotted as grey curves. The mean across this ensemble of ACFs (\(\langle\mathscr{R}\rangle\)) is plotted as a cyan dashed curve and \(1\sigma\) spread about the mean is indicated by the vertical bars. The quasi-universal form of the ACF (\(\mathscr{R} = e^{-\ell_x/\lambda_x}\)) is plotted as a deep-pink curve. \textit{Bottom right}: ``Box-and-whisker'' plot where each element (horizontal line) of a box-and-whisker from bottom to top indicates minimum, \(10^\text{th}\) percentile, median, \(90^\text{th}\) percentile, and maximum, at the respective lag. Equivalently, the green-shaded region within each box indicates the middle 80\% of the distribution. A reference exponential function is plotted in deep pink. See text for further details.}
\label{fig:vK_sim}
\end{figure*}

The top left panel of Fig. \ref{fig:vK_sim} shows \(R(\ell_x)\) computed as described above for all the averaging domains, and the bottom left panel shows the ACFs normalized by their respective zero-lag values. Domains associated with ARs are identified as those having a maximum magnetic field magnitude above 500 G \citep[e.g.,][]{vanDriel-Gesztelyi2015LRSP} and the corresponding ACFs are depicted in orange, while QS ACFs are shown in blue. The large variability of the unnormalized ACFs (\(R\)), both in the associated energies and decorrelation with lag, is highlighted in the top left panel. The transformation in the bottom left panel to relatively similar forms following normalization (\(R'\)) is striking. The decay of the \(R'\) correlations indicates a form familiar from experiments, observations, and numerical simulations of turbulence \citep[e.g.,][]{Batchelor1953book,matthaeus1982JGR,Yeung1989JFM}. ARs have systematically longer decay scales than the QS, consistent with \cite{Abramenko2024SoPh} and examined further below. The majority of ACFs have decayed to small values (\(\lesssim 0.1\)) by \(\ell_x=15\) Mm. The ACFs \(R(\ell_y)\) have a very similar behavior (not shown).

\begin{figure}  
\centering
\includegraphics[width=0.4\textwidth]{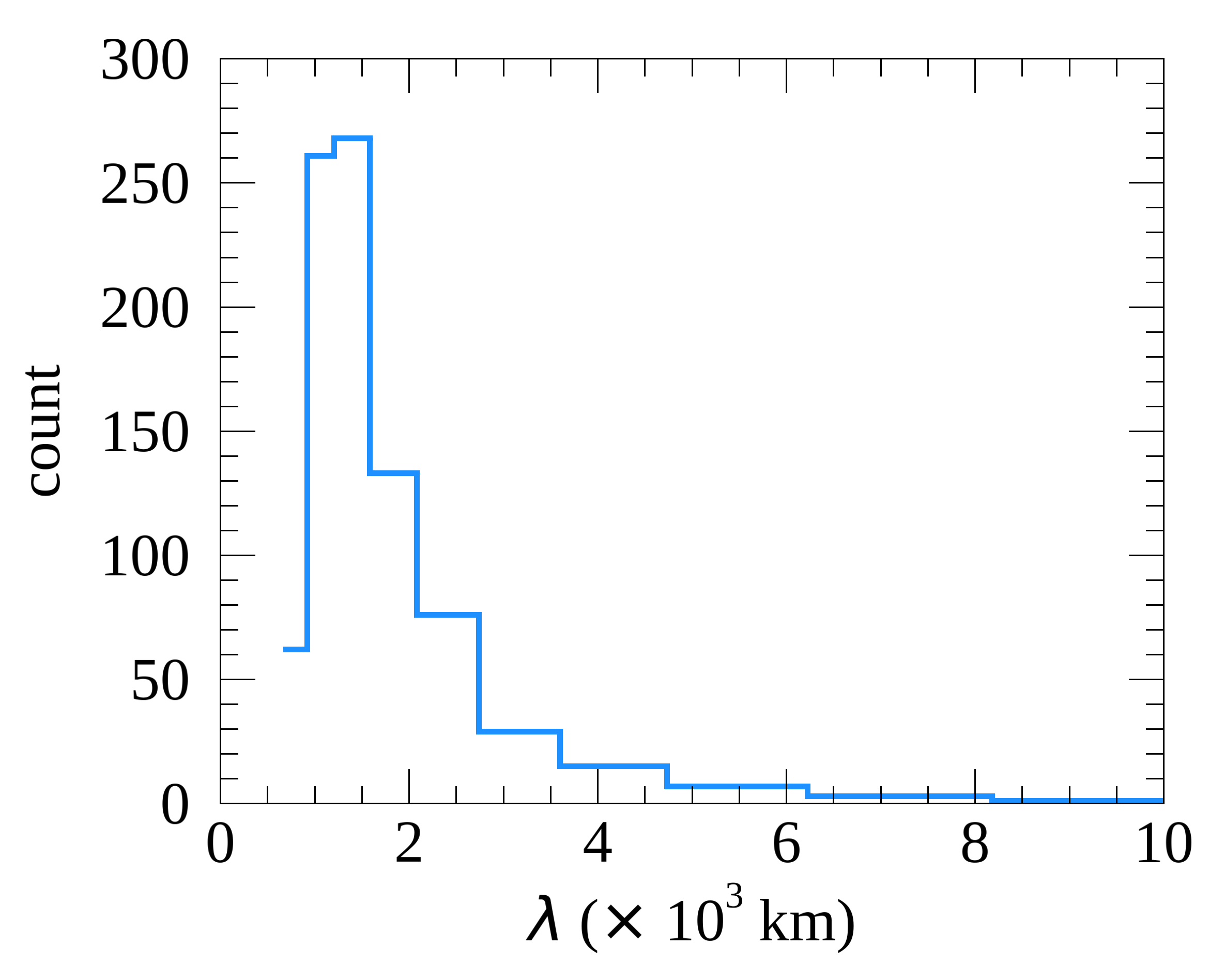}
\includegraphics[width=0.45\textwidth]{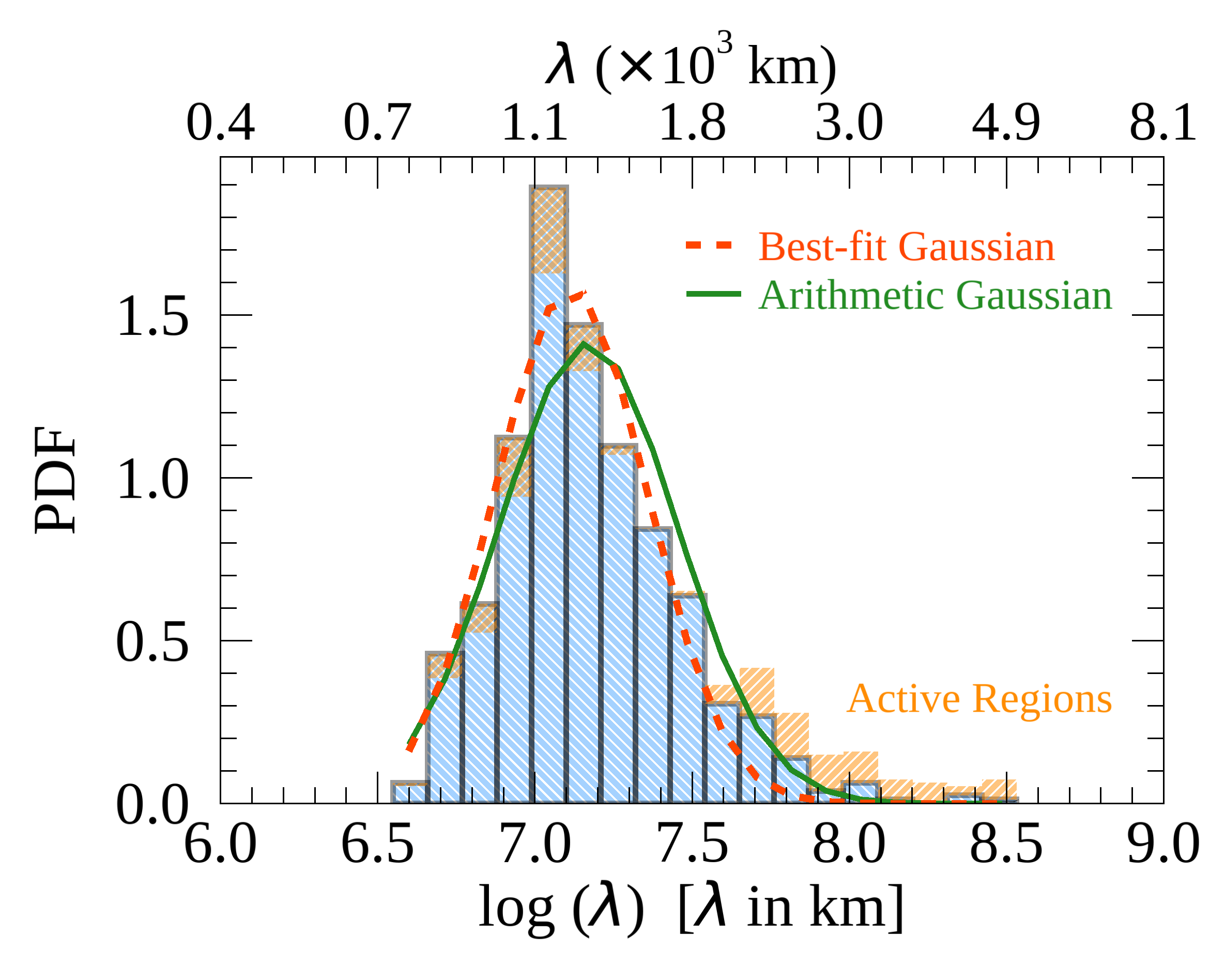}
        \caption{\textit{Top}: Histogram of correlation lengths \(\lambda\) (that combines \(\lambda_x\) and \(\lambda_y\)) of magnetic fluctuations, computed using Eq. \eqref{eq:lambda1}. \textit{Bottom}: ``Stacked'' bar plot showing the probability density function (PDF) of \(\log \lambda\), with selected values of \(\lambda\) indicated on top axis. Blue bars with thick outlines show the PDF with \(\lambda\) values associated with ARs removed. The orange region above or below the top of each blue bar indicates the modification to the PDF when ARs are included in the ensemble. Blue and orange bars have stripes oriented at different angles. Dashed red curve shows a best-fit Gaussian to the blue PDF of \(\log \lambda\). Green curve shows an ``arithmetic'' Gaussian , i.e., one with the same mean and standard deviation as the blue PDF. See text for more details.}
\label{fig:pdf}
\end{figure}

\vspace{2mm}
\noindent \textit{II. Correlation Lengths and von K\'arm\'an--Howarth Similarity.} The ACFs computed above generate an ensemble for evaluating the vK-H similarity hypothesis. Examining Eq. \eqref{eq:vkH} we note that a first step towards evaluating the universal form \(\mathscr{R}(\ell/\lambda)\) has already been completed by normalizing each ACF by its respective zero-lag value \(R(0)=\langle b^2\rangle\), to produce \(R'(\ell_x)\) \citep[see also][]{Roy2021ApJ}. To proceed we require the correlation length of each ACF. We first address ACFs in which oscillatory behavior at small lags precludes unambiguous estimation of \(\lambda\) by identifying the lag \(\ell_e\) where an ACF \(R'\) is nearest to \(1/e\) and checking if the ACF reaches a value of 0.3 at any lag \textit{below} \(\ell_e\); such cases are discarded from subsequent analyses. Next, \(\lambda_x\) is computed from Eq. \eqref{eq:lambda1} with the upper limit of integration taken to be \(3\ell_{0.3}\), where \(\ell_{0.3}\) is the lag at which the ACF first reaches a value smaller than 0.3.\footnote{In the cases where \(3\ell_{0.3}\) is greater than the maximum lag evaluated (i.e., \(1/5^\text{th}\) of the ``length'' of the averaging domain, as described in Sec \ref{sec:res}.I) the upper limit of integration is this maximum lag.} This upper limit is prescribed since we are interested in \textit{local} correlations associated with turbulent dynamics, and we wish to exclude long-range correlations or periodicities that presumably originate in very large-scale structures like coronal holes or the solar dynamo \citep[e.g.,][]{Mursula1996JGR,Wang2025arXiv}. The distribution of computed correlation lengths is shown in the bottom panel of Fig. \ref{fig:magnetogram} and in Fig. \ref{fig:pdf}, and is discussed further below. A computation of correlation lengths based on the \(1/e\) method was also performed, finding similar distributions (not shown; however, see Appendix \ref{sec:app2}). 

Proceeding with the vK-H similarity analysis, we rescale the lags (i.e., the arguments of a correlation function \(R'\)) associated with each ACF to \(\ell_x/\lambda_x\), where \(\lambda_x\) is the respective correlation length. The resulting functions \(\mathscr{R}(\ell_x/\lambda_x)\) are shown as grey curves in the top right panel of Fig. \ref{fig:vK_sim}, as a function of the rescaled lags. One immediately notices the collapse to an apparent quasi-universal form and the reduced statistical spread in the ensemble. The deep-pink curve shows an  exponentially decaying function \(e^{-\ell_x/\lambda_x}\) for reference, which evidently passes through the approximate middle of the ensemble. For a more precise comparison we compute a mean \(\mathscr{R}\) across the ensemble: First, each ACF \(\mathscr{R}\) is linearly interpolated to a regular grid of 100 (rescaled) lag values in the interval \([0,10]\); if the largest lag \((\ell_x/\lambda_x)_\text{max}\) for an ACF is smaller than 10 then the ACF is interpolated to the regular grid in the interval \([0,(\ell_x/\lambda_x)_\text{max}]\) and ``padded'' with zeros in the interval \(((\ell_x/\lambda_x)_\text{max},10]\) \citep{matthaeus1982JGR}. The mean across the ensemble of ACFs is then computed at each lag on the regular grid, yielding the function \(\langle\mathscr{R}\rangle\), plotted as a dashed cyan curve. Vertical ``error bars'' on this curve indicate the \(1\sigma\) spread about the mean where \(\sigma\) denotes a standard deviation. The overlap between \(\langle\mathscr{R}\rangle\) and the exponential function is significant.

To further quantify the statistical spread of the rescaled ACFs \(\mathscr{R}\), the bottom right panel of Fig. \ref{fig:vK_sim} shows a ``box and whisker'' plot of the ensemble: Each green box contains the middle \(80\)\% of the ACFs at the respective (rescaled) lag; bottom and top ends of a green region indicate the 10th and 90th percentile values respectively while the black horizontal line inside the green box indicates the median of the distribution.\footnote{This rendering of the box plot differs from the more common ``quartile plot'', where the box contains only the middle 50\% of a distribution with bottom and top edges indicating \(25^\text{th}\) and \(75^\text{th}\) percentiles, respectively.} The bottom and top ``whiskers'' (above and below the box) indicate the minimum and maximum values of the distribution at the respective lag. Once again, the collapse to the quasi-universal exponential form is evident, reinforcing the vK-H similarity hypothesis. Appendix \ref{sec:app2} describes further analysis of the statistical collapse to a quasi-universal form, and also presents results based on correlation lengths computed using the \(1/e\) method, with similar conclusions as above.

\vspace{2mm}
\noindent \textit{III. Probability Distribution of Correlation Lengths.} Fig. \ref{fig:pdf} shows a histogram (top panel) and a PDF (bottom panel) of the correlation lengths computed above [using Eq. \eqref{eq:lambda1}]. Here \(\lambda\) includes both \(\lambda_x\) and \(\lambda_y\) to obtain a larger statistical sample; their separate PDFs (not shown) are quite similar. The histogram, which is shown with a linear scale on the horizontal axis, contains  the telltale one-sided tail towards large values of \(\lambda\), a suggestion of a lognormal PDF \citep{Limpert2001Biosci}. The most probable value is  \(\sim 1500\) km. The PDF plot is shown for \(\log \lambda\) on the bottom axis, with selected values of \(\lambda\) displayed on the top axis. Values of \(\lambda\) above 5000 km are excluded from the PDF due to insufficient counts (see top panel). The  plot contains two PDFs shown as a ``stacked'' bar plot: blue bars with thick outlines show the PDF after excluding \(\lambda\) values associated with ARs, while the orange region above or below the top of each blue bar indicates the modification to the PDF when ARs are included in the ensemble. Note that for \(\log \lambda \lesssim 7.5\) ARs tend to ``pull down'' the PDF while above this value they tend to increase the PDF, indicating that large values of \(\lambda\) are associated with ARs, as also seen in Fig. \ref{fig:vK_sim}. This is consistent with the recent study of ARs by \cite{Abramenko2024SoPh}.

A best-fit of the three-parameter Gaussian function \(A_0 \exp{[-(X-A_1 )^2 /(2A_2 ^2)]} \) to the blue PDF (with ARs removed) is computed, where \(X\equiv\log \lambda\). This nonlinear least-squares fit yields \(A_0=1.6,~A_1=7.1~\text{and}~A_2=0.2\).\footnote{The IDL function \href{https://www.nv5geospatialsoftware.com/docs/GAUSSFIT.html}{{\fontfamily{cmtt}\selectfont
gaussfit.pro}} is used. \(A_0, A_1,\) and \(A_2\) are the height, center, and standard deviation of the Gaussian, respectively, with \(1\sigma\) error estimates 0.09, 0.02, and 0.02.} This function is plotted as a dashed-red curve, while the green curve shows an ``arithmetic'' Gaussian, i.e., one with the same mean and standard deviation as the blue PDF. The figure indicates that the PDF of correlation lengths (excluding ARs) is approximately described by a lognormal distribution. The corresponding fit to the PDF that includes ARs is less good (not shown), suggesting that turbulent fluctuations in the quiet Sun may constitute a different population from those in active regions. A more detailed and systematic study of the goodness of fits of the PDF \citep[e.g.,][]{Ruiz2014SoPh} will require a larger sample,  utilizing a large number of full-disk magnetograms that cover an extended time period, and will be taken on in future work. We note here that the lognormal distribution has been shown to describe other elements of the photosphere, including the sizes of  supergranules \citep{Noori2019AdvSpacRes} and distributions of magnetic flux \citep{Abramenko2005ApJ}.

Following the characterization scheme suggested by \cite{Limpert2001Biosci} for lognormal distributions (which they call \textit{multiplicative normal distributions}), we report the \textit{geometric} mean of \(\lambda\):\\ \(\lambda^* \equiv \sqrt[n]{\prod_{i=1}^{n} \lambda_i}= 1433\) km, where \(\prod\) denotes a product and the index \(i\) runs over the full distribution of \(n\) values of \(\lambda\) contained in the blue PDF. Note that \(\lambda^*\) is equivalent to \(\exp \left({\frac{1}{n}\sum_{i=1}^{n} \log\lambda_i}\right)\), which is the median of the lognormally distributed \(\lambda\) \citep{Limpert2001Biosci}. The multiplicative standard deviation is \(\sigma^* \equiv\exp\sigma =1.46\) where \(\sigma\) is the (usual) standard deviation of \(\log\lambda\), again computed from the distribution contained in the blue PDF. Then the 68.3\% interval of confidence is \(1433  \timesdiv 1.46\) km or [982,~2092] km. Here \(\timesdiv\) implies times/divide, corresponding to plus/minus for the established sign \(\pm\). The 68.3\% interval of confidence expressed here corresponds to the familiar interval \(\mu\pm s\) for a normal distribution with mean \(\mu\) and standard deviation \(s\). See \cite{Limpert2001Biosci} for more details.  

We also report the mean and standard deviation of the lognormally distributed \(\lambda\) in terms of the parameters \(A_1\) and \(A_2\) of the best-fit Gaussian computed above \citep{Aitchison1957book,Abramenko2005ApJ}: \(\text{mean}~(\lambda)\equiv \exp\left (A_1 + A_2^2/2 \right)=1237\) km and \(\text{standard deviation}~ (\lambda)\equiv \exp (2A_1+2A_2^2)-\exp(2A_1 +A_2^2)=250\) km.

\begin{figure}  
\centering
\includegraphics[width=0.45\textwidth]{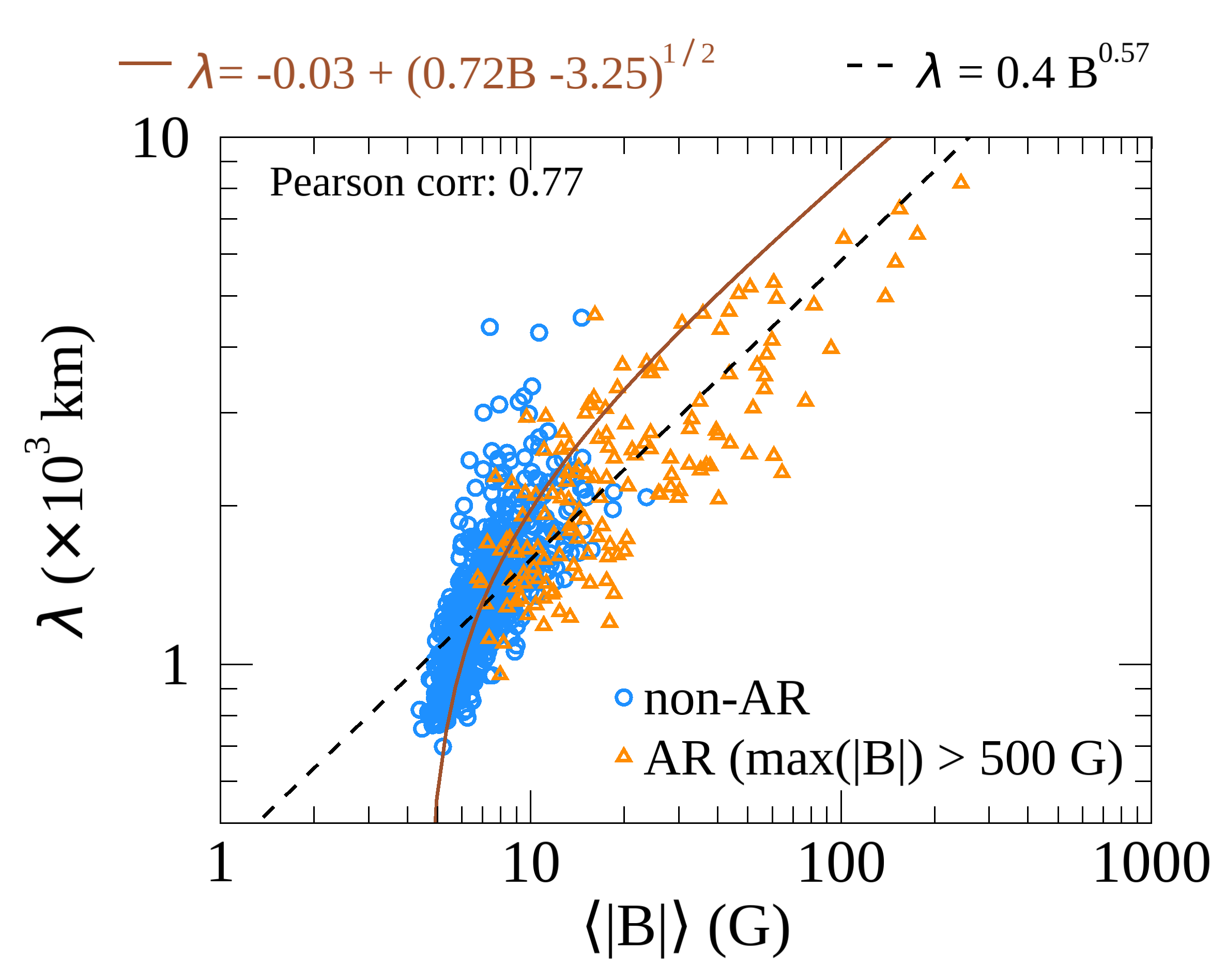}
        \caption{Scatterplot of average magnitude of magnetic field within each averaging domain versus correlation length computed within the respective domain. Averaging domains with active regions are identified by requiring the maximum value of the magnetic field within a domain to be greater than 500 G.}
\label{fig:scatter}
\end{figure}

\vspace{2mm}
\noindent \textit{IV. Correlation of \(\lambda\) with Mean Magnetic Field Strength.} We have seen that magnetic fluctuations in ARs have larger correlation lengths than the QS, as has also been noted in other recent work \citep{Abramenko2024SoPh}. This could potentially indicate differences in turbulence dynamics occuring in QS and ARs. At the same time, since ARs have larger magnetic field magnitudes than the QS, our results also suggest a correlation between the large-scale magnetic field \(B\) and \(\lambda\). To probe this further we show in Fig. \ref{fig:scatter} a scatter plot of \(\lambda\) (combining \(\xhat\) and \(\yhat\) components) as a function of the mean magnitude of the magnetic field \(\langle|B|\rangle\), where averaging is performed over the same domains as those for the respective ACFs (Sec. 4.I). A significant correlation is apparent, with a computed Pearson correlation coefficient \citep[e.g.,][]{Hoel1960Stats} of 0.77. QS and ARs are represented with distinct symbols, reinforcing the previous findings of large \(\lambda\) in the latter. Even in the QS (blue circles), a positive correlation between \(\lambda\) and magnetic field is evident. 

This relationship is further quantified by fitting a power law\footnote{The IDL function \href{https://www.nv5geospatialsoftware.com/docs/linfit.html}{{\fontfamily{cmtt}\selectfont linfit.pro}} was used to fit a straight line of the form \(\log\lambda = \log A_0 + A_1\log B\) by minimizing the \(\chi\)-squared error statistic, to obtain a power law of the form \(\lambda=A_0B^{A_1}\). The result was \(\log A_0=-0.84\pm0.03\) and \(A_1=0.57\pm0.01\).}, 
finding \(\lambda=0.4 B^{0.57}\), where \(\langle|B|\rangle\) has been denoted simply as \(B\) to avoid visual clutter. 
We also fit a \(2^\text{nd}\)-order polynomial\footnote{The IDL function \href{https://www.nv5geospatialsoftware.com/docs/robust_poly_fit.html}{{\fontfamily{cmtt}\selectfont robust\_poly\_fit.pro}} was used to perform an outlier-resistant least-square polynomial fit to the quadratic equation \(B=A_0 + A_1\lambda +A_2\lambda^2\), finding \(A_0=4.50,~A_1=0.08,~\text{and}~A_2=1.38\). The quadratic formula was used to find the real solution for \(\lambda\) from the above equation.}, finding \(\lambda=-0.03+(0.72B-3.25)^{0.5}\).
The polynomial fit appears to describe the QS data better, which is expected on account of the outlier-resistant fitting approach. The power law provides a simpler function that roughly describes the relationship between \(B\) and \(\lambda\). We do not attempt to interpret further the nature of the observed correlation between \(\lambda\) and \(\langle|B|\rangle\) here. However, it is important to note that the provided empirical fits can be useful in turbulence transport models \citep[TTMs; e.g.,][]{breech2008turbulence,usmanov2018} of the solar wind; specifically, a rough estimate of the spatial distribution of \(\lambda\) at the photospheric boundary may be obtained simply from a map of the magnetic field. This novel approach can provide data-driven boundary conditions for TTMs \citep{Huang2023ApJ}, an improvement over the typically-used heuristic, spatially-uniform values for \(\lambda\) at the near-Sun boundary \citep[e.g.,][]{vanderholst2014ApJ,usmanov2018,zank2018ApJ}.

\vspace{2mm}
\noindent \textit{V. Mosaic of Correlation Lengths over the Photosphere.} The bottom panel of Fig. \ref{fig:magnetogram} presents a mosaic of the correlation lengths distributed over the photosphere, which gives a visual representation of their spatial inhomogeneity, and their relationship with magnetic field strength and active regions.\footnote{To obtain a less ``noisy'' visual, this plot is based on a more coarse-grained computation of the ACF \(R(\ell_x)\) than the one used for previous results; here the averaging domain is \(300\times 300\) pixels.} Evidently the correlation scale can change by a factor of five within a \(\sim10\degree\) span of latitude/longitude. This change can be even more dramatic in the case of larger ARs than the ones present in the current magnetogram, where \(\lambda\) can be as large as 30 Mm \citep{Abramenko2024SoPh}.

\section{Conclusions and Discussion}\label{sec:Disc}

Compared to numerous and extensive investigations of turbulence in the solar wind, turbulence in the solar photosphere has received relatively limited attention \citep{Petrovay2001SSR,Rincon2018LRSP}. This is partly due to the lack of in-situ observations in the region; however, current remotely-observed magnetograms offer sufficient resolution to probe the larger scales of photospheric turbulence \citep[e.g.,][]{Abramenko2001AstronRep,McAteer2010AdvSpacRes,Abramenko2024SoPh}. The current era of space physics research, defined by the Parker Solar Probe's exploration of the corona (down to a height of less than \(9~R_\odot\) from the solar surface) is an opportune time for exploiting these magnetogram datasets to better understand the solar origin and injection of turbulence into the heliosphere and to establish  more firmly the connections between photospheric and coronal turbulence. In this paper we have used such a magnetogram dataset to examine some properties of autocorrelation functions of magnetic fluctuations and the associated correlation lengths, quantities that are of fundamental importance to models of turbulence and energetic particle transport in the heliosphere \citep[e.g.,][]{breech2008turbulence,Engelbrecht2022SSR}. We summarize and discuss our main findings below.

\begin{table*}
\centering
\begin{tabular}{||c c c c c c c||} 
  (\(\perp\) mag. corr. len. &  Photosphere & Chromosphere \& T.R. & Coronal base & Young solar wind & Near-Earth &  \\ 
  in km) & \(1~R_\odot\) & 1.001 -- 1.01 \(R_\odot\) & 1.01 -- 1.3 \(R_\odot\) & 0.1 -- 0.6 AU & 1 AU & 2 -- 40 AU \\
 \hline\hline
 Corr. length &  1500\textsuperscript{~(a)} & 1500 -- 5000\textsuperscript{~(d)} & 3500 -- 8000\textsuperscript{~(e)} & \(10^4~\dash~ 0.7\times 10^6\)\textsuperscript{~(f)} & \(10^6\)\textsuperscript{~(g)} & \(10^6 \dash 10^7\)\textsuperscript{~(h)}
 \\ 
  Corr. length in a coronal hole & 1000\textsuperscript{~(b)}  & 
  \\
 Corr. length in ARs & \(8000~ \dash~40,000\)\textsuperscript{~(c)} & \\ 
 \hline
\end{tabular}
\caption{\(\perp\) correlation lengths of magnetic turbulence in the photosphere, and evolution with distance from the Sun. For comparison, the length scales associated with granules and supergranules are 500-2000 km and 30,000 km, respectively \citep[][and references within]{Rincon2018LRSP}. All lengths are in km and are approximate values; see sources (below) for details on uncertainties and statistical spreads. Here ``\(\perp\) mag. corr. len.'' stands for perpendicular magnetic correlation length, defined relative to a mean magnetic field direction, and T.R. stands for transition region. Note that the top row refers to a general correlation length, not necessarily associated with a particular type of photospheric structure. Sources of values in the table are as follows. (a) Chhiber et al. (2025, present work). (b) \protect{\citet{Abramenko2013ApJ}}. (c) \protect{\citet{Abramenko2024SoPh}}; Chhiber et al. (2025, present work). (d) \protect{\citet{Bailey2025ApJ}}. (e) \protect{\citet{Sharma2023NatAst,Bailey2025ApJ,Hahn2025ApJ_corrlen}}. (f) \protect{\citet{Ruiz2014SoPh,chhiber2021ApJ_psp,cuesta2022ApJL}}. (g) \protect{\citet{Ruiz2014SoPh,cuesta2022ApJL}}. (h) \protect{\citet{smith2001JGR,breech2008turbulence,Ruiz2014SoPh,adhikari2017ApJ,cuesta2022ApJL}}.}
\label{table:1}
\end{table*}

\noindent\textit{I.} The magnetogram data we examined was from a period of low solar activity, and included both ``quiet Sun'' and active regions. The mean value of the computed correlation lengths (\(\lambda\)) was \(\sim 1500\) km, although the distribution had a large variability with an approximately lognormal PDF. This mean value is close to the correlation length computed by \cite{Abramenko2013ApJ} in a coronal-hole region, and is comparable to the typical scale of solar granulation (see Table \ref{table:1}). It has been noted that the phenomenological scales of turbulent convection in the Sun appear to be associated with granular rather than supergranular scales on the surface \citep{Rincon2018LRSP}. A value near 1000 km also appears to be consistent with recent estimates of the correlation length in the chromosphere and low corona \citep[see Table \ref{table:1};][]{Sharma2023NatAst,Bailey2025ApJ,Hahn2025ApJ_corrlen}. Furthermore, \cite{Abramenko2013ApJ} point out that analysis of higher resolution magnetograms (e.g., from the \textit{Hinode} mission) lead to lower estimates of \(\lambda\), suggesting that the present values may be an upper bound. We found that active regions had systematically larger \(\lambda\) than the quiet Sun, consistent with the recent study by \cite{Abramenko2024SoPh}.

\noindent \textit{II.} The correlation length is an important parameter in solar-wind turbulence transport models (TTMs), requiring specification at the photospheric boundary; while some TTMs have specified \(\lambda\sim 100\) km at the photosphere \citep[e.g.,][]{cranmer2007ApJS}, others have chosen values closer to the supergranulation scale \citep[\(\sim 30,000\) km; e.g.,][]{verdini2007apj,usmanov2018}. Our discussion above motivates the need for a reassessment of the proper photospheric boundary conditions for TTMs. One future direction is offered by our finding of a significant positive correlation between the average magnetic field strength in the photosphere and the correlation length of the turbulence: our empirical fits, while crude, provide a simple and computationally inexpensive way to estimate a spatially-varying distribution of \(\lambda\) across the photosphere from low-resolution maps of the photospheric magnetic field, including the synoptic maps used in global solar wind models \citep[e.g.,][]{riley2014SoPh}.

\noindent \textit{III.} We confirmed the validity of the von K\'arm\'an--Howarth similarity hypothesis for photospheric magnetic turbulence. The correlation functions of highly dissimilar regions in the photosphere \citep[see also][]{Abramenko2013ApJ,Abramenko2024SoPh} collapse to a quasi-universal exponential form following a rescaling by their respective energies and correlation lengths. The exponential nature of the correlation function itself has deep roots in terms of diagnosing the state of self-organization of a system \citep{Watkins2016SSR}. We note, however, that while the exponential decorrelation describes the large-scale end of the inertial range and the energy-containing (and larger) scales, the smaller-scale behavior (near the dissipation scale) may be different \citep[][]{taylor1938ProcRSL,Bandyopadhyay2020ApJ}. Of course, probing the small-scale end of the inertial range in a reliable manner is challenging with current magnetogram resolutions.

The vK-H hypothesis is the basis for phenomenological treatments of turbulence decay of the form
\begin{equation}
    \text{decay rate} \propto \frac{(\text{fluctuation amplitude})^3}{\text{similarity scale}},
\end{equation}
where the similarity scale can be associated with the correlation scale \citep[e.g.,][]{breech2008turbulence,chhiber2021ApJ_psp}. This type of formalism is widely used in solar wind models to account for plasma heating \citep[e.g.,][]{usmanov2011solar,vanderholst2014ApJ,downs2016ApJ}, and our results, together with recent work validating this hypothesis in the solar wind at 1 AU \citep{Roy2021ApJ,Roy2022PRE}, provide significant conceptual support for the use of this approach in astrophysical systems. 

We carry out a brief and crude calculation here to estimate the turbulent heating rate in the photosphere based on our present dataset. This heating rate per unit mass is taken to be \(Q_T/\rho\sim \alpha Z^3/\lambda\) \citep[see, e.g.,][]{hossain1995PhFl,usmanov2018}, where the K\'arm\'an-Taylor constant \(\alpha\) is assumed to be unity, \(Z^2 = v^2 + b_A^2\) is twice the sum of the average energy densities in velocity and magnetic fluctuations, with magnetic fluctuations expressed in Alfv\'en speed units: \(b_A^2=b^2/(4\pi\rho)\). Assuming equipartition between velocity and magnetic fluctuation energies so that \(v^2=b_A^2\) \citep[as is the case for Alfv\'en waves;][]{bittencourt2004plasma,Jess2009Sci}, and further assuming equipartition between the three components of the magnetic variance, we estimate \(Q_T/\rho\sim [6~b_{r}^2/ (4\pi\rho) ]^{3/2}/\lambda\), where \(b_r\) is the radial component. Next, we assume that \(b_r\) is equal to the line-of-sight component \(b_\text{los}\). This is estimated as the standard deviation of the line-of-sight component of the photospheric magnetic field, first computed within the averaging subdomains described in Sec. \ref{sec:res}.\textit{I} and then averaged over all the subdomain values, to find \(b_\text{los}\sim 20\) G. Note that this fluctuation amplitude is equivalent to \(\sim0.4 ~\text{km}~\text{s}^{-1}\) in Alfv\'en units \citep[cf.][]{cranmer2007ApJS,verdini2007apj,Jess2009Sci}, with a photospheric proton-mass density of \(\rho=1.67\times 10^{-7} ~\text{g}~\text{cm}^{-3}\) \citep[e.g.,][]{priest1982book}. Using \(\lambda\sim1500\) km, the above expression yields \(Q_T/\rho\sim 3\times10^5 ~\text{erg}~\text{s}^{-1}~\text{gm}^{-1}\), which is comparable to the lower end of the photospheric heating rates used in the model of \cite{cranmer2007ApJS}. Further consideration of this important topic is left to future work.

\noindent \textit{IV.} Our results offer new insights on the origin and evolution of turbulence in the heliosphere.  The vK-H similarity collapse supports the notion that the solar photosphere is in a turbulent state \citep{Petrovay2001SSR}, which in turn reflects on possible turbulent properties in the solar interior \citep[][]{Rincon2018LRSP}. Table \ref{table:1} lists approximate values of the correlation length at different distances from the Sun, starting from \(\sim 1500\) km at the photosphere, increasing to \(\sim 5000\dash 8000\) km at the coronal base, \(\sim 10^4 \dash 10^6\) through the young solar wind up to 1 AU, and reaching \(\sim 10^7\) km by 40 AU. This increase is likely a combined effect of geometrical expansion effects \citep[e.g.,][]{Hollweg1986JGR} and further in-situ development of turbulence \citep[e.g.,][]{matthaeus1998JGR,bruno2013LRSP,deforest2016ApJ828,chhiber2018apjl}. Note that the top row of Table \ref{table:1} refers to a general correlation length, not necessarily associated with a particular type of photospheric structure. Future studies that focus on the radial evolution of correlation lengths associated with specific photospheric sources may ``fill out'' the second and third rows of the Table.

The approximately lognormal PDF of correlation lengths in the photosphere suggests that non-linear multiplicative processes are occurring at the source of the solar wind, with further evolution of the PDF via in-situ processes taking it closer to a lognormal form as the solar wind expands into the inner heliosphere \citep[see][]{Ruiz2014SoPh}. These results have implications for our understanding of ``\(1/f\) noise'' in frequency spectra across the heliosphere \citep[e.g.,][]{Wang2024SoPh}. 
In particular a well-known generic path to producing $1/f$ spectra is the superposition of powerlaw signals with a scale-invariant weighting. 
But it has also been shown that a lognormal distribution can closely approximate a scale-invariant distribution over a number of decades in frequency \citep{Montroll82}.
Therefore the 
presence of lognormality in photospheric correlation scales supports the idea that $1/f$ signals might originate in the solar dynamo
\citep[][and references within]{Wang2024SoPh}.

We end by remarking on possible avenues for future work that will be of interest. In contrast to the line-of-sight component of the magnetic field examined here (which yielded transverse correlations), vector magnetograms \citep{Hoeksema2014SoPh} can be used to study the longitudinal correlation function of the \(b_\theta,b_\phi\) components, which may reveal anisotropy introduced by the mean magnetic field \citep[e.g.,][]{dasso2005ApJ,Wang2022ApJ,Wang2024ApJ}. Another obvious extension would be to analyze higher-resolution magnetograms (\textit{Hinode}, DKIST), to explore further the possibility of finding smaller correlation scales \citep[see][]{Abramenko2013ApJ}. We note here that we have performed preliminary computations of correlation scales using relatively low-resolution synoptic maps of the magnetic field provided by HMI, and the resulting values are not significantly different from the ones presented here. Our estimates of the turbulent correlation scale could aid in constraining diffusion scales in models of magnetic flux transport on the solar surface \citep{Jiang2014SSR}. Finally, analysis of a large number of magnetograms, along with categorization by solar activity levels, would make the types of statistical analyses performed here more robust and provide insights on trends with solar activity. These magnetogram-based studies can be complemented by those employing EUV images \citep[e.g.,][]{Georgoulis2005SoPh} and novel remote observations of the corona from NASA's PUNCH mission \citep{DeForest2025arXiv_PUNCH}.







----------------------------------------\\

\section*{Acknowledgements}
R.C. acknowledges useful discussions on the subject of this paper with Cooper Downs, Riddhi Bandyopadhyay, and Mahendra Verma. This research was supported by NASA under the Living With a Star (LWS) Science program grant 80NSSC22K1020 and utilized resources provided by the \href{https://dspoc.org/}{Delaware Space Observation Center (DSpOC)} numerical facility. 
W.H.M. and S.R. are partially supported by NSF grant PHY-2108834. 
The study used the \href{https://www.lmsal.com/solarsoft/}{IDL SolarSoft library} \citep{Freeland2012ASCL}, the \href{https://github.com/wlandsman/IDLAstro}{IDL Astronomy User's library}, and the \href{https://github.com/idl-coyote}{IDL Coyote library}.

\section*{Data Availability}
SDO/HMI data are publicly available at Stanford University's \href{http://jsoc.stanford.edu/}{Joint Science Operations Center (JSOC)}.
 



\bibliographystyle{mnras}





\appendix

\section{Illustration of ACF Calculation}\label{sec:app0}

\begin{figure}  
\centering
\includegraphics[width=0.5\textwidth]{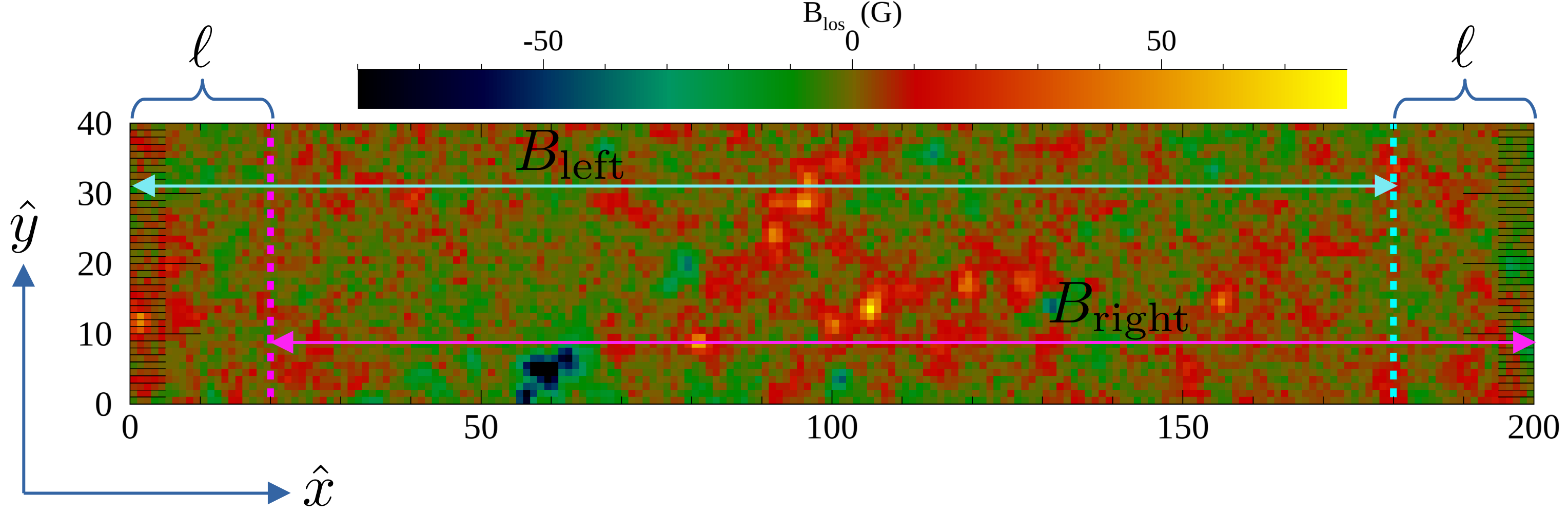} \\
\vspace{4mm}
\includegraphics[width=0.4\textwidth]{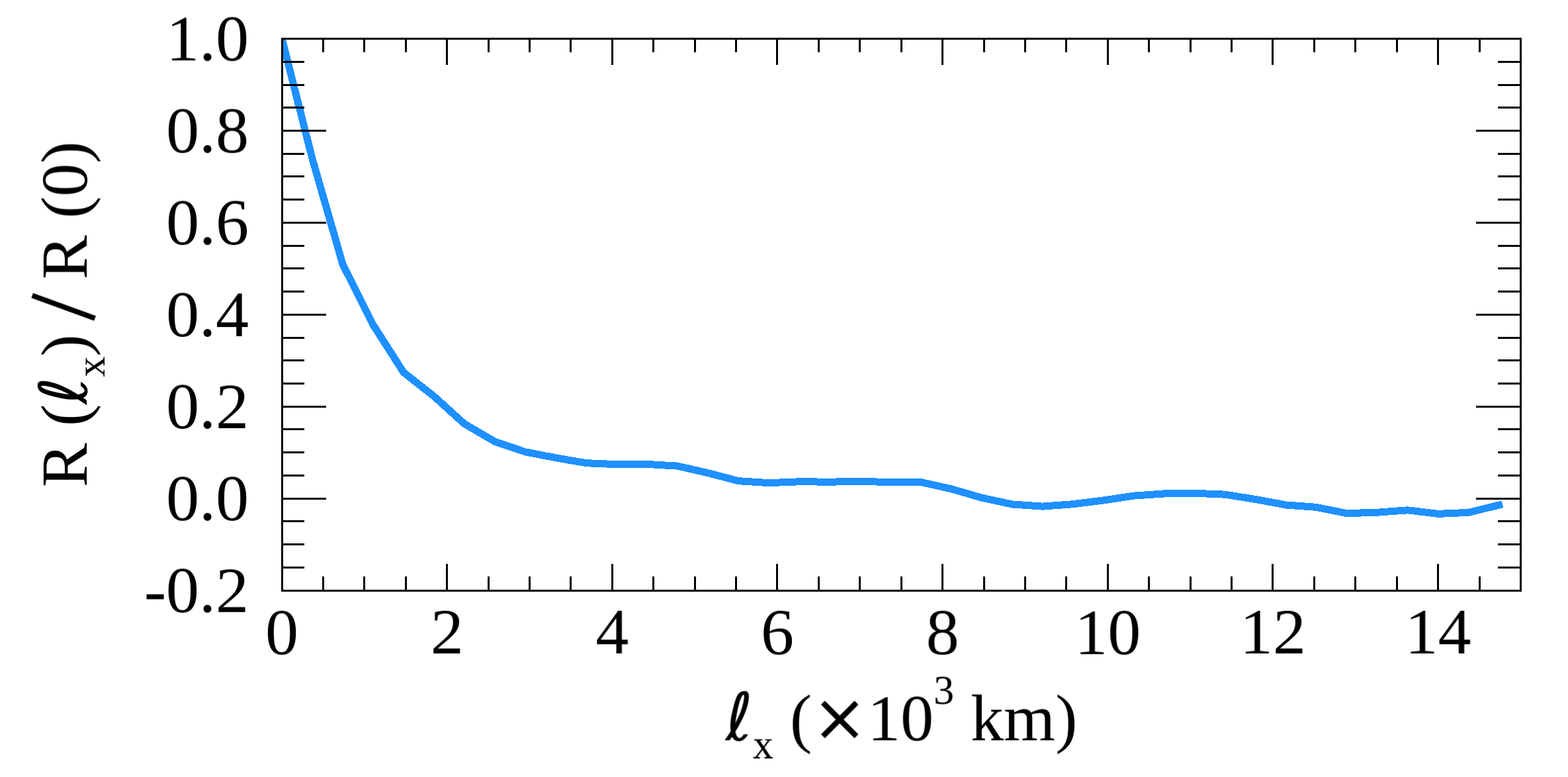}
        \caption{\textit{Top}: Example averaging domain in which the ACF \(R(\ell_x\equiv\ell)\) is computed, with dimensions \(200\times 40\) pixels, as indicated with tick marks. Averaging domains for \(B_\text{left}\) and \(B_\text{right}\) are indicated with cyan and magenta colored lines, respectively. See text for more details. \textit{Bottom}: ACF computed in this domain.}
\label{fig:illust}
\end{figure}

This appendix illustrates the procedure used to calculate ACFs from the photospheric data. The top panel of Fig. \ref{fig:illust} shows an example averaging domain of size \(200\times40\) pixels, over which the ACF \(R(\ell_x)\) is computed (see Sec \ref{sec:res}.I). The bottom left corner of this domain is at 77\degree~ Carrington longitude and 9\degree~ latitude. The los magnetic field in this domain is denoted as a 2-dimensional (2D) array \(B[L_x,L_y]\), with the first and second array dimensions denoting \(\hat{\bm{x}}\) and \(\hat{\bm{y}}\) directions, respectively. In our case  \(L_x=200\) and \(L_y=40\). Adapting the \cite{Blackman1958} method for computing ACFs, for a given lag of \(\ell_x\equiv\ell\) pixels we define two sets of arrays:
\begin{equation}
    B_\text{left} \equiv B[1:L_x-\ell,\ast], \hspace{4mm}  
     B_\text{right}\equiv B[\ell+1:L_x,\ast],
\end{equation}
where the arguments within square brackets indicate a range of indices (starting at 1) and `\(\ast\)' indicates all elements, in the usual way. Then we have [see Eq. \eqref{eq:R2}] 
\begin{equation}
    R(\ell_x) = \langle B_\text{left} B_\text{right}\rangle - \langle B_\text{left} \rangle \langle B_\text{right} \rangle,
\end{equation}
where \(B_\text{left} B_\text{right}\) indicates an element-wise product of two equal-sized arrays, and \(\langle M\rangle\) denotes an arithmetic mean over the entire 2D array \(M\) \citep[cf.][]{Roy2021ApJ}. The averaging domains associated with \(B_\text{left}\) and \(B_\text{right}\) are schematically shown in the top panel of Fig. \ref{fig:illust}. The bottom panel shows the ACF computed in this domain, normalized to its zero-lag value.

\section{Correlation Length in relation to the Magnetic Energy Spectrum}\label{sec:app1}

\begin{figure}  
\centering
\includegraphics[width=0.4\textwidth]{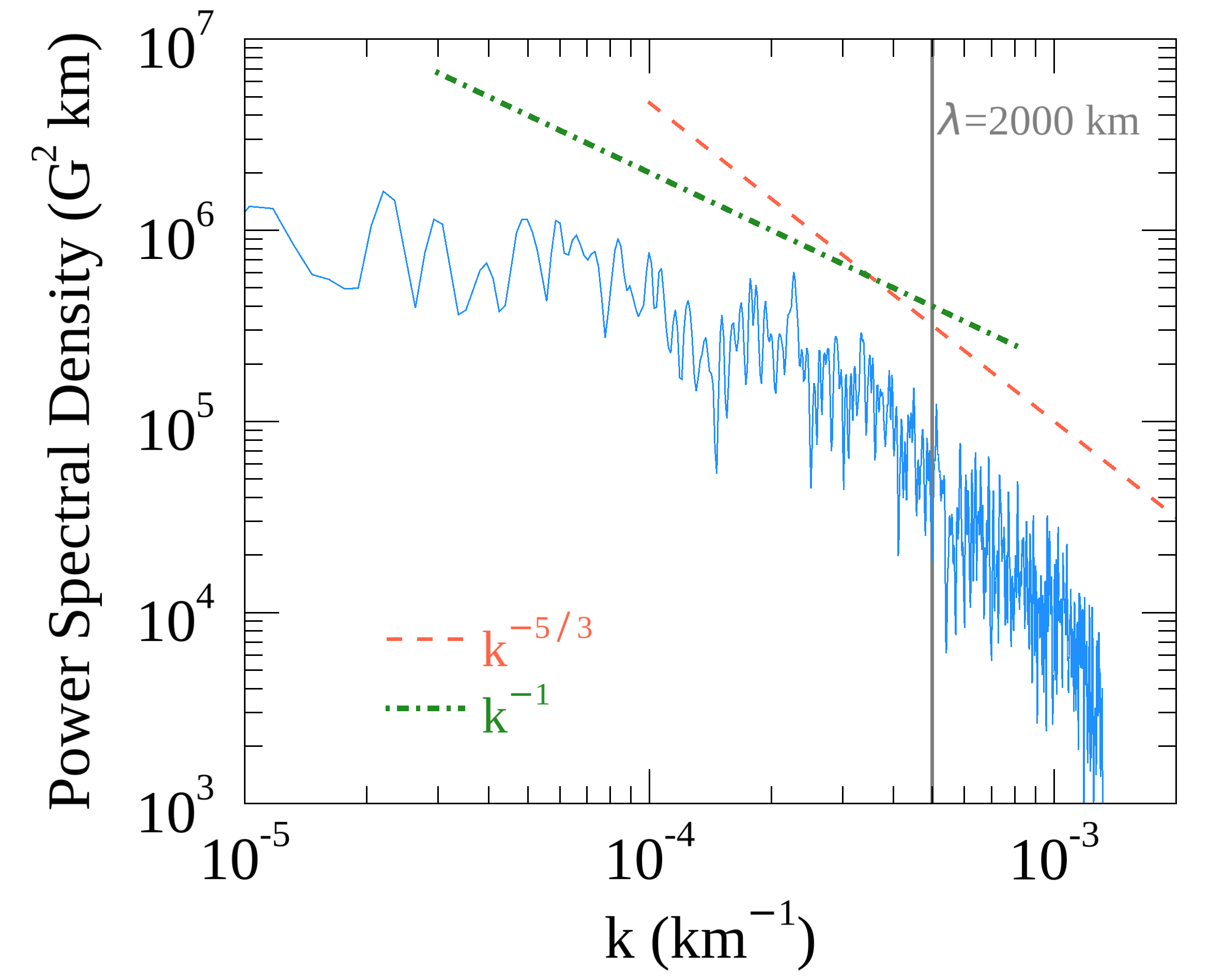}
        \caption{Blue curve shows power spectral density of line-of-sight magnetic field in the photosphere as a function of wavenumber \(k\), computed as described in the text. Red dashed and green dash-dotted lines indicate reference \(k^{-5/3}\) and \(k^{-1}\) spectra, respectively. The inertial-range \(\sim -5/3^\text{rd}\) spectrum transitions into the shallower ``\(1/k\)'' range at scales slightly larger than the approximate correlation length on the quiet Sun, with the latter indicated with a vertical grey line at \(k=1/2000 ~\text{km}^{-1}\).}
\label{fig:spectrum}
\end{figure}

As noted in Sec. \ref{sec:backg}, the correlation length is associated with the low-wavenumber edge of the inertial-range energy spectrum, at which energy is injected from large-scales to smaller scales where the turbulent cascade occurs \citep[e.g.,][]{matthaeus1982JGR,pope2000book}. In Fig. \ref{fig:spectrum} we show a sample  energy spectrum of the photospheric (los) magnetic field in which the aforementioned association is evident. This power spectral density (PSD) is computed for a horizontal ``slice'' of \(B_\text{los}\) at the center of the image shown in the top panel of Fig. \ref{fig:magnetogram}, over the entire horizontal range of data. The Fast Fourier Transform (FFT) approach is used to compute the power spectrum as a function of wavenumber \(k\)\footnote{We use the IDL function \href{https://www.nv5geospatialsoftware.com/docs/FFT_PowerSpectrum.html}{{\fontfamily{cmtt}\selectfont FFT\_PowerSpectrum.pro}} with Tukey-filter smoothing. The smoothing window has a width of 0.002, expressed as a fraction of the number of points. Note that wavenumber \(k\) corresponds to a spatial scale \(\sim1/k\).},
and the PSD is then computed by dividing this spectrum by \(\Delta k\), where \(\Delta k\) is the difference between adjacent wavenumbers. The figure indicates that the HMI full-disk magnetogram used here resolves the low-\(k\) part of a Kolmogorov-type turbulent inertial range with \(\sim k^{-5/3}\) \citep[similar photospheric spectra have been shown by, e.g.,][]{Abramenko2001AstronRep,McAteer2010AdvSpacRes}. The spectrum ``bends over'' to a shallower \(\sim1/k\) range \citep{Nakagawa1974ApJ,Wang2024SoPh} at a scale of roughly \(5000\) km. At even smaller \(k\) the PSD becomes almost flat. Note that the PSD can be computed directly as the Fourier transform of the correlation function \citep{matthaeus1982JGR}; such an approach may yield further insight into the relationship between the spectral ``break'' wavenumber and the correlation scale. Further examination of photospheric spectra lies outside the scope of the present study and will be taken on in future work.

\section{Additional Analyses of Correlation Functions}\label{sec:app2}

\begin{figure}  
\centering
\includegraphics[width=0.45\textwidth]{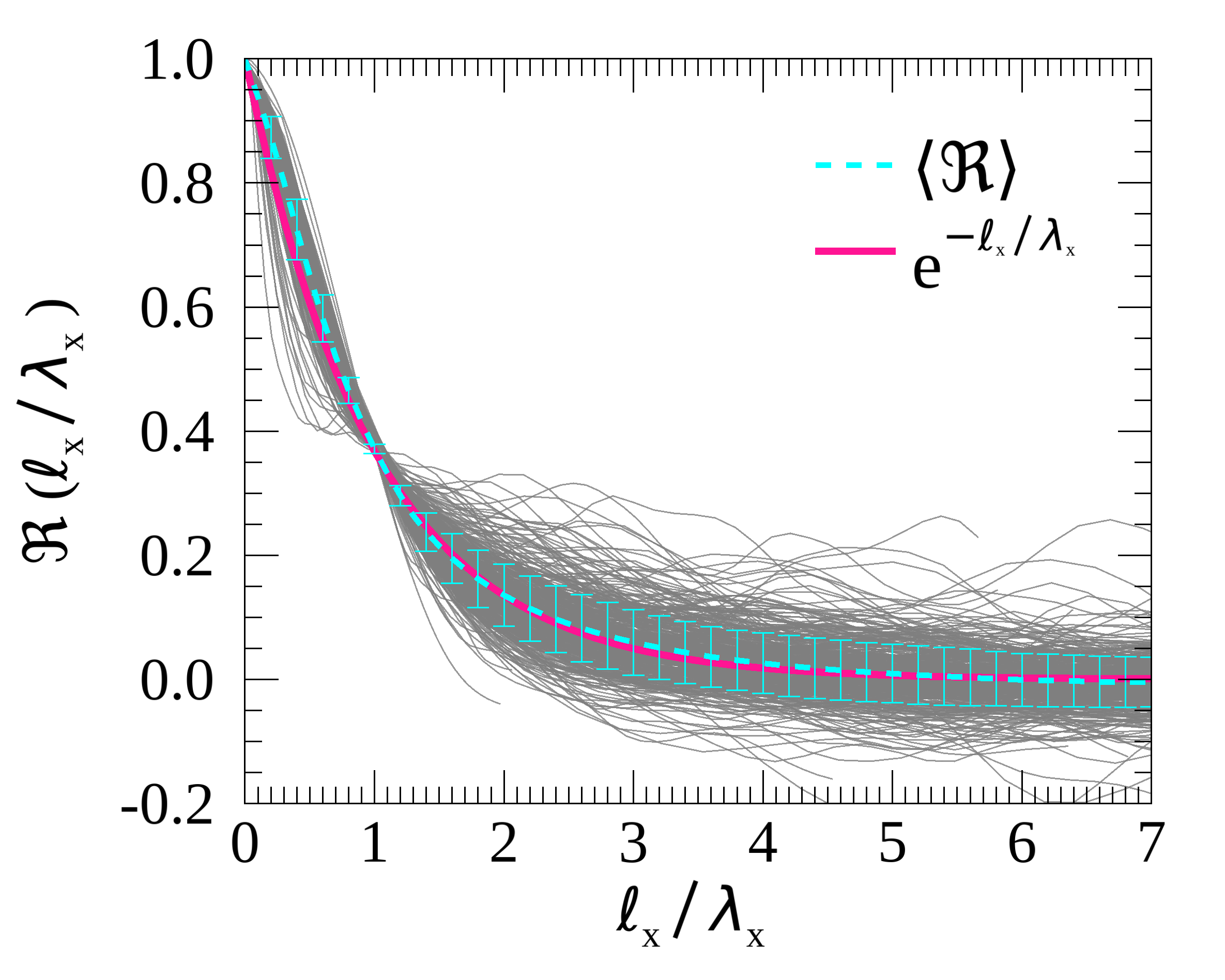}
        \caption{Ensemble of ACFs with lags scaled by respective correlation lengths, with the latter computed using the ``\(1/e\)'' method. See text for details. The remaining description follows from that of Fig. \ref{fig:vK_sim}, top right panel.}
\label{fig:vK_sim2}
\end{figure}

In this section we describe additional analyses of the ACFs and their vK-H similarity behavior. In Fig. \ref{fig:vK_sim2} we show the rescaled ACFs, starting from the same ensemble of ACFs as in the bottom left panel of Fig. \ref{fig:vK_sim}, but in this instance using the \(1/e\) method to compute the respective correlation lengths, in contrast to Eq. \eqref{eq:lambda1}. These correlation lengths were computed as follows: (i) ACFs that display oscillatory behavior at small lags were discarded from the ensemble as described in Sec. \ref{sec:res}.II; (ii) For the remaining ACFs, we identified the smallest lag \(\ell_{0.3}\) where an ACF is below or equal to a value of 0.3; (iii) Within the range \([0,\ell_{0.3}]\), we used linear interpolation to estimate the lag where the respective ACF attains a value of \(1/e\), thus obtaining the correlation length. It is clear that this method produces, by construction, rescaled ACFs \(\mathscr{R} (\ell_x/\lambda_x)\) that are identically equal to \(1/e\) at \(\ell_x/\lambda_x=1\); this leads to the ``pinch'' seen in the ACFs in Fig. \ref{fig:vK_sim2}. Note that the distribution (not shown) of correlation lengths computed using the above method is very similar to that obtained using Eq. \eqref{eq:lambda1} and discussed in the main body of the paper.

\begin{figure}  
\centering
\includegraphics[width=0.45\textwidth]{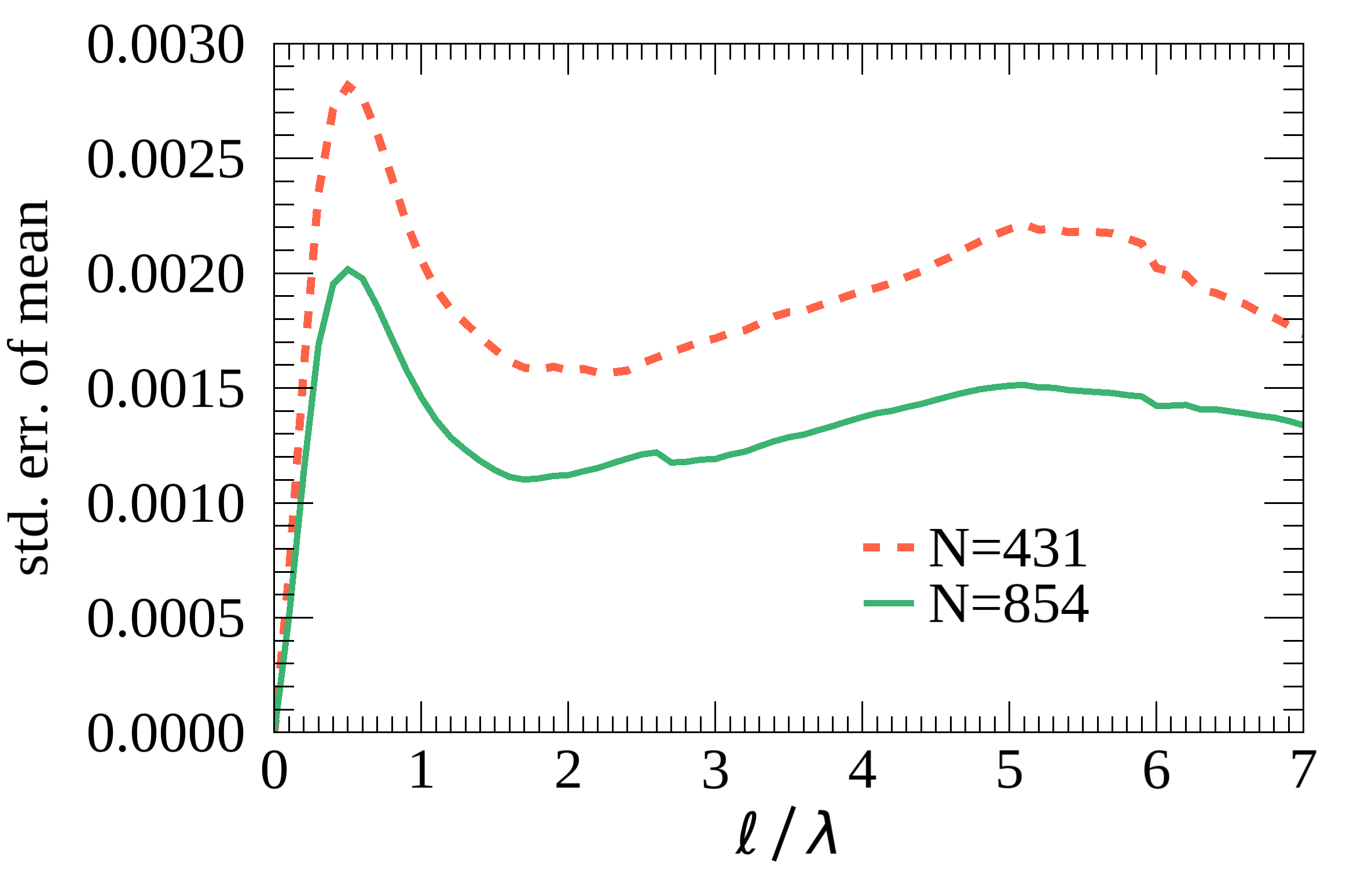}
        \caption{Standard error of the mean of the ensemble of rescaled ACFs reduces with increasing sample size \(N\), suggesting convergence of the ensemble of ACFs to a quasi-universal form. See text for details.}
\label{fig:vK_sim3}
\end{figure}

Next, we test the convergence of the rescaled ACFs to a quasi-universal form as the sample size of the ACF ensemble is increased. For this purpose we evaluate the standard error of the mean \citep[e.g.,][]{Hoel1960Stats} for the mean ACF \(\langle\mathscr{R}\rangle\) that was shown in the top right panel of Fig. \ref{fig:vK_sim}. This error of the mean is computed at each lag using the formula \(\sigma/\sqrt{N}\) where \(\sigma\) is the standard deviation of the ensemble of ACFs at a given lag and \(N\) is the sample size. For the ensemble shown in the middle panel of Fig. \eqref{fig:vK_sim} \(N=431\), and the resulting standard error of the mean is plotted as the dashed-red curve in Fig. \ref{fig:vK_sim3}. To increase the sample size we next include the ACFs with lags in the \(\yhat\) direction, assuming that they come from the same population as the \(\xhat\) ACFs, noting the very-similar nature of their statistical properties. This approximately doubles the population to \(N=854\), and the resulting standard error of the mean is shown as the green curve in Fig. \ref{fig:vK_sim3}. The decrease in this error with increasing sample size suggests a corresponding systematic convergence of the ensemble of ACFs to a quasi-universal form \citep[see also][]{Roy2021ApJ}, lending further support to the von K\'arm\'an--Howarth similarity hypothesis for photospheric magnetic turbulence.

%

\bsp	
\label{lastpage}
\end{document}